%% file: main.tex
\documentclass[sigconf]{acmart}
\AtBeginDocument{%
  }

\acmSubmissionID{96}

\copyrightyear{2025}
\acmYear{2025}
\setcopyright{cc}
\setcctype{by}
\acmConference[UIST '25]{The 38th Annual ACM Symposium on User Interface Software and Technology}{September 28-October 1, 2025}{Busan, Republic of Korea}
\acmBooktitle{The 38th Annual ACM Symposium on User Interface Software and Technology (UIST '25), September 28-October 1, 2025, Busan, Republic of Korea}
\acmDOI{10.1145/3746059.3747690}
\acmISBN{979-8-4007-2037-6/2025/09}

\usepackage{wrapfig}
\usepackage{color,soul}

\input{macros}

\begin{document}

\title{Squidgets: Sketch-based Widget Design for Scene Manipulation}

\author{Joonho Kim}
\orcid{0000-0002-2521-9813}
\affiliation{%
  \institution{University of Toronto}
  \city{Toronto}
\country{Canada}}
\email{joonho@dgp.toronto.edu}

\author{Fanny Chevalier}
\orcid{0000-0002-5585-7971}
\affiliation{%
  \institution{University of Toronto}
  \city{Toronto}
\country{Canada}}
\email{fanny@dgp.toronto.edu}

\author{Karan Singh}
\orcid{}
\affiliation{%
  \institution{University of Toronto}
  \city{Toronto}
\country{Canada}}
\email{karan@dgp.toronto.edu}

\renewcommand{\shortauthors}{Kim et al.}

\begin{abstract}
  People naturally sketch strokes over graphical scenes to convey scene changes. We propose automatically interpreting these strokes to execute scene changes with squidgets ({\em sketch-widgets}), a novel sketch-based UI framework for direct scene manipulation. Squidgets are motivated by the observation that curves resulting from visually abstracting scene elements provide natural handles for the direct manipulation of scene parameters. Additional curves can be defined by users to author custom handles associated with scene attributes. Users manipulate a scene by simply drawing strokes, that are partially matched against scene curves to select a squidget and interactively control scene parameters associated with the squidget. We present an implementation of squidgets within the 3D animation system {\it Maya}, showing 2D/3D stroke input to manipulate 2D/3D scenes. We report on a controlled experiment evaluating squidgets on 2D object translation and deformation tasks, and a broader informal study on squidget creation and manipulation.
\end{abstract}

\ccsdesc[500]{Human-centered computing~Interaction design}

\keywords{sketching, graphics, 3D scenes, interactive design}

\begin{teaserfigure}
  \centering
  \includegraphics[width=0.98\textwidth]{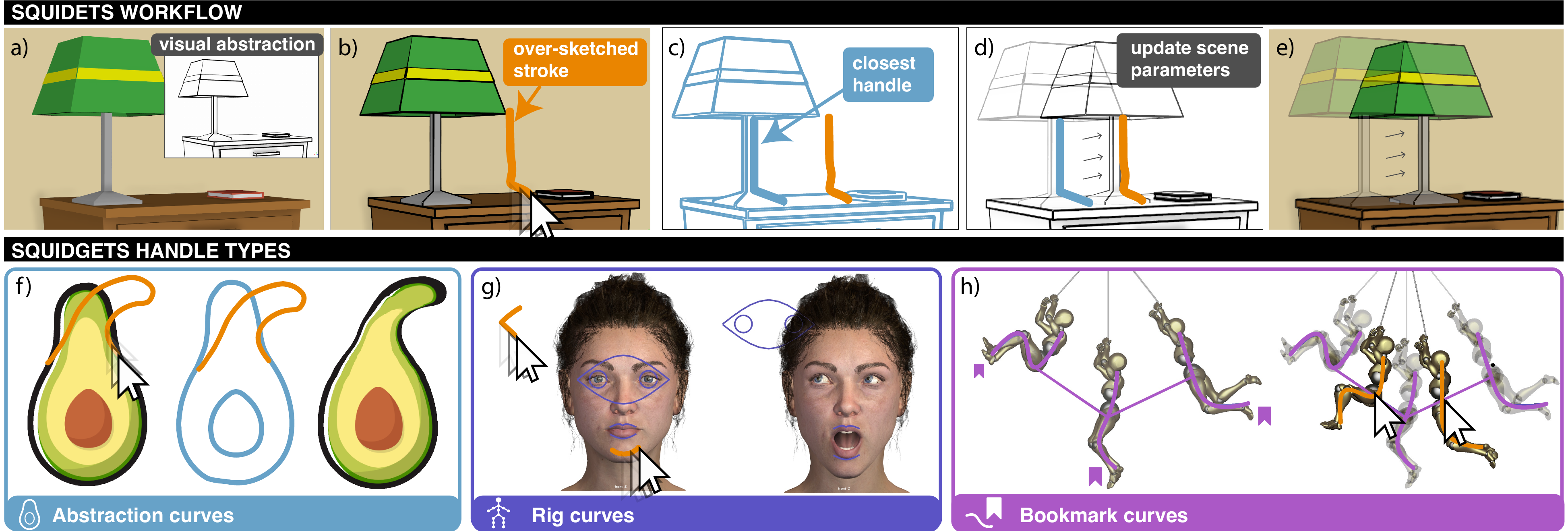}
  \caption{Humans naturally communicate desired scene changes by over-sketching (a,b). These sketched \stroke{strokes}, when aligned with visual \abscurve{abstraction curves} of a scene (c), can effectively express user intended changes to scene attributes (d,e). Squidgets or `sketched widgets' enable interactive manipulation of scene attributes via scene curves (f-h), such as the avocado shape attributes deformed to match its silhouette \abscurve{abstraction curve} (f). All scene curves can homogeneously function as squidgets including pre-defined \rigcurve{rig curves} like the gaze/jaw control curves on a 3D facial rig (g), and user drawn \bookmarkcurve{bookmark curves} that capture  attribute configurations of objects, like the line-of-action curves drawn to pose an animated character (h).
  }
\Description{The top row of images show from left to right: a) a 3D lamp asset with an inset of the lamps outlines, b) the same image with a user stroke next to the lamp in the shape of the lamp post outline, c) abstraction curves of the lamp and what curve the stroke matches with, d-e) the lamp moving to the curve.  Underneath are 3 images from left to right: a) an avocado illustration with a stroke drawn bending the avocado head, b) 2 strokes drawn to pose a neutral face rig with the eyes and mouth rig controls, c) bookmark curves drawn to overlay a swinging character motion.}
\label{fig:teaser}
\end{teaserfigure}

\received{9 April 2025}
\received[revised]{14 July 2025}
\received[accepted]{30 July 2025}

\maketitle

\input{sections/1_intro}

\input{sections/2_relatedwork}

\input{sections/3_squidgets}

\input{sections/4_implement_framework}

\input{sections/5_applications}
\input{sections/6_study}

\input{sections/7_conclusion}

\begin{acks}
\add{
We would like to thank: our user-study participants for providing insightful feedback on our tool; Jenny Oh for help organizing our user study; Damien Masson for helping create data figures; Zhecheng Wang for helping with the curve matching algorithm.
}

\add{
\textit{Park Environment} in \autoref{fig:perf_all} and \textit{Cartoon Bedroom Set} furniture in \autoref{fig:teaser} were modeled by @anim\_matt.
\textit{Valley Girl} characters in \autoref{fig:teaser} and \autoref{fig:lady} and \textit{Angela} in \autoref{fig:animators} were modeled by \copyright{Chris Landreth}.
\textit{Squirrels Rig} in \autoref{fig:perf_all}: Rig or Material used with permission (\copyright{Animation Mentor 2022}). No endorsement or sponsorship by Animation Mentor. Downloaded at \url{www.animationmentor.com/free-maya-rig/}.
VR environments were rendered in \copyright{Gravity Sketch}.
\textit{FRUIT ICONS} for the userstudy and in \autoref{fig:teaser} were designed by NicoDigitalStore.
This research was supported by NSERC.
}
\end{acks}

\bibliographystyle{ACM-Reference-Format}
\bibliography{references}


\end{document}

%% file: macros.tex
\usepackage{colortbl}   
\usepackage{xcolor}
\usepackage{xspace}
\usepackage{subcaption}
\usepackage{enumitem}

\newcommand{\add}[1]{\textcolor{black}{#1}}
\newcommand{\removed}[1]{}

\newcommand{\redacted}[1]{\textbf{REDACTED}}

\definecolor{c_stroke}{RGB}{234, 133, 0}
\definecolor{c_abscurve}{RGB}{103,162,200}
\definecolor{c_rigcurve}{RGB}{91,83,197}
\definecolor{c_bookmarkcurve}{RGB}{171,88,198}

\definecolor{c_1stroke}{RGB}{220, 38, 127}
\definecolor{c_selectdrag}{RGB}{0, 11, 214}
\definecolor{c_selectpre}{RGB}{0, 158, 115}
\definecolor{c_selectmanip}{RGB}{0, 217, 159}

\setul{0.5ex}{0.3ex}

\newcommand{\stroke}[1]{\setulcolor{c_stroke}\ul{#1}}
\newcommand{\abscurve}[1]{\setulcolor{c_abscurve}\ul{#1}}
\newcommand{\rigcurve}[1]{\setulcolor{c_rigcurve}\ul{#1}}
\newcommand{\bookmarkcurve}[1]{\setulcolor{c_bookmarkcurve}\ul{#1}}

\newcommand{\baseline}{\emph{baseline}\xspace}
\newcommand{\onestroke}{\emph{1-stroke}\xspace}
\newcommand{\selectdrag}{\emph{select$\rightarrow$drag}\xspace}
\newcommand{\selectmanipulate}{\emph{select$\rightarrow$manipulate}\xspace}

%% file: sections/1_intro.tex
\section{Introduction}

\begin{figure}[tbp]
\centering \includegraphics[width=\linewidth]{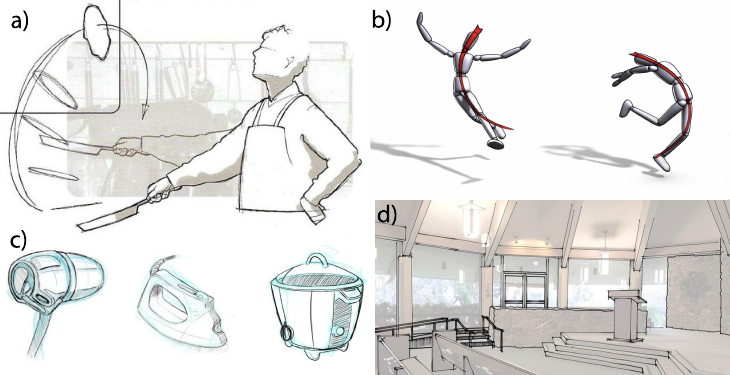}
\caption{Traditional sketch stroke usage to convey aspects of animation: motion paths ({\footnotesize \copyright{K. Eissen, R, Steur, BIS Publishers}})~\cite{Eissen08} (a),
line-of-action \add{({\footnotesize \copyright{} M. Guay})}~\cite{lineofaction} (b),
design (over-sketched ideation {\footnotesize \copyright{R. Arora}})~\cite{arora2017sketchsoup} (c),
 spatial overlays \add{({\footnotesize \copyright{E. Mikiten}})} \cite{archlayers} (d).}
\Description{A 4 panel figure with illustrations showing various stroke abstractions for drawing and design. From top left to bottom right, (a) shows a cook flipping a pancake and the motion path of the pancake as a stroke curve. (b) shows the line of action for a cartoon cat punching and a cartoon cat receiving the punch. (c) shows sketch-abstractions for designing an earbud, iron, and rice cooker. (d) shows an auditorium overlaid with sketch lines to show the corners and edges of objects in the room.}

\label{fig:squidget-motivation}
\end{figure}

Sketching is a traditionally established tool for visual communication (\autoref{fig:squidget-motivation}). 
Digital sketching is now also ubiquitous in interactive design and graphical content creation. 
User control of scene objects and attributes/parameters in such applications however is typically a disparate combination of traditional UI components, pre-defined widgets, and stroke gestures~\cite{sketchwidgets08}. 
While a number of seminal sketch-based systems for content creation span over half a century~\cite{sketchpad,zeleznik,teddy,ilovesketch,xu2014true2form}, their operational interface has largely relied on pre-defined stroke gestures and traditional UI elements.

Pen and touch stroke-based interfaces are also increasingly used to control general computing applications. 
A vein of research has thus adapted the interaction of UI menus/buttons~\cite{crossy}, sliders~\cite{tsandilas2015sketchsliders}, and 3D widgets~\cite{sketchwidgets08}, from a point-and-click to a stroke-friendly design.
Inspired by recent research~\cite{juxtaform} using sketchy-renderings of objects for interactive visualization and exploration, we aim to exploit such renderings for stroke-based scene interaction.

We thus present \textbf{squidgets} ({\em sketch-widgets}) as a novel stroke-based UI framework for direct scene manipulation (\autoref{fig:teaser}).
Our work uses the insight that graphical scenes inherently possess implicit, in-situ handles for scene manipulation.
People naturally sketch a variety of strokes (\autoref{fig:squidget-motivation}) over perceived scene curves
to visually indicate desired scene changes (\autoref{fig:teaser}(a-e)). Practically, scene curves are either visually imagined \abscurve{abstraction curves} (like the silhouette of the avocado in \autoref{fig:teaser}(f)) or explicitly modeled scene objects (like the gaze-controller \rigcurve{rig curve} in \autoref{fig:teaser}(g) or the line-of-action \bookmarkcurve{bookmark curves} in \autoref{fig:teaser}(h)).
We aim to capture this natural interaction with scene curves using squidgets. Conceptually, this is an inverse rendering problem: {\em compute a minimal change to the current scene attribute values that will result in a manipulated scene, in which some scene curve (real or imagined) visually matches the over-sketched stroke} (\autoref{fig:teaser}(a-f)). 

The benefits of solving such a problem are twofold. 
First, over-sketching is natural, direct, and in-situ for 2D (\autoref{fig:teaser}(f)) or 3D (\autoref{fig:teaser}(g,h)) applications. 
Second, no explicit gestures, widgets or UI components for the scene need to be defined or learned by the user. 
There are however, three non-trivial and potentially ambiguous aspects to solving this ambitious problem: (1) inferring the user-imagined \abscurve{abstraction curves} of a graphical scene; (2) discerning what part of this curve abstraction to associate with the over-sketched stroke; and (3) computing changes to the values of a set of scene attributes that result in a manipulated scene whose associated visual abstraction best matches the over-sketched stroke.

The squidgets framework allows each of these problems to be explored, constrained, and addressed independently.
(1) Visual abstraction of scenes as curves have been extensively studied in Non-Photorealistic Rendering (NPR) literature \cite{gooch2001non}. 
Such abstractions can be defined using occluding and suggestive contours in a scene \cite{hertzmann1999introduction} (\autoref{sec:squidgets}), or using image-based approaches to differential rendering \cite{image2lines}. 
(2) We define a novel curve similarity metric (\autoref{sec:implement}) that is able to match the stroke to (partial) curves using a perceptual mix of corner, spatial, and shape proximity. A user stroke can thus be matched to a curve segment in the scene (whether explicitly modeled as a scene object, or inferred as a visual scene abstraction). 
(3) Scene attributes that deform the associated curve segment can then be sampled around their present values to find a resulting curve edit that best matches the user stroke.

Increasingly, artists are hand-crafting custom UI layouts using \rigcurve{rig curves} for in-situ scene manipulation \cite{kim21} (\autoref{fig:teaser}(g)). Such curves being explicit scene objects are homogeneously handled by our framework. Inspired by artistic constructs like the line-of-action for posing animated characters \cite{lineofaction} (\autoref{fig:squidget-motivation}(b)), we further support the rapid creation of \bookmarkcurve{bookmark curves} that are associated with a given configuration of scene attributes (\autoref{fig:teaser}(h)) within our interaction framework. 
Note that \abscurve{abstraction}/\rigcurve{rig}/\bookmarkcurve{bookmark} curves are simply a categorization of all scene curves (real and imagined) that can serve as interaction handles in our framework as persistent scene objects (\rigcurve{rig curves} and \bookmarkcurve{bookmark curves}) explicitly define both a visual scene abstraction (problem 1), and the set of scene attributes associated with the curves (problem 2). Further, since these curves and their scene attributes, are related by simple invertible (often linear) functions, finding attribute values to optimally match an over-sketched stroke is simple (problem 3).

The squidgets framework thus addresses the stroke-based
manipulation of scene attributes via {\bf all visually perceived curves} in the scene (imagined \abscurve{abstraction curves}, or explicitly modeled scene curves: \rigcurve{rig curves} and  \bookmarkcurve{bookmark curves}).

We provide an overview of related work (\autoref{sec:related-work}), followed
by details of our squidgets framework (\autoref{sec:squidgets}).
\autoref{sec:implement} presents our approach to stroke matching, attribute inference and other implementation details of a squidgets interaction prototype built within the modeling and animation system {\em Maya}. \autoref{sec:applications} presents several squidgets usage scenarios.
We discuss the outcomes of a user study (\autoref{sec:study}) followed by a discussion of overall insights, limitations, and avenues for future work on squidgets (\autoref{sec:discussion}).

%% file: sections/2_relatedwork.tex
\section{Related Work}
\label{sec:related-work}




The squidgets framework touches upon many areas of graphics and HCI research that we roughly classify into four themes as follows. 

\begin{figure}[!htbp]
    \centering \includegraphics[width=\linewidth]{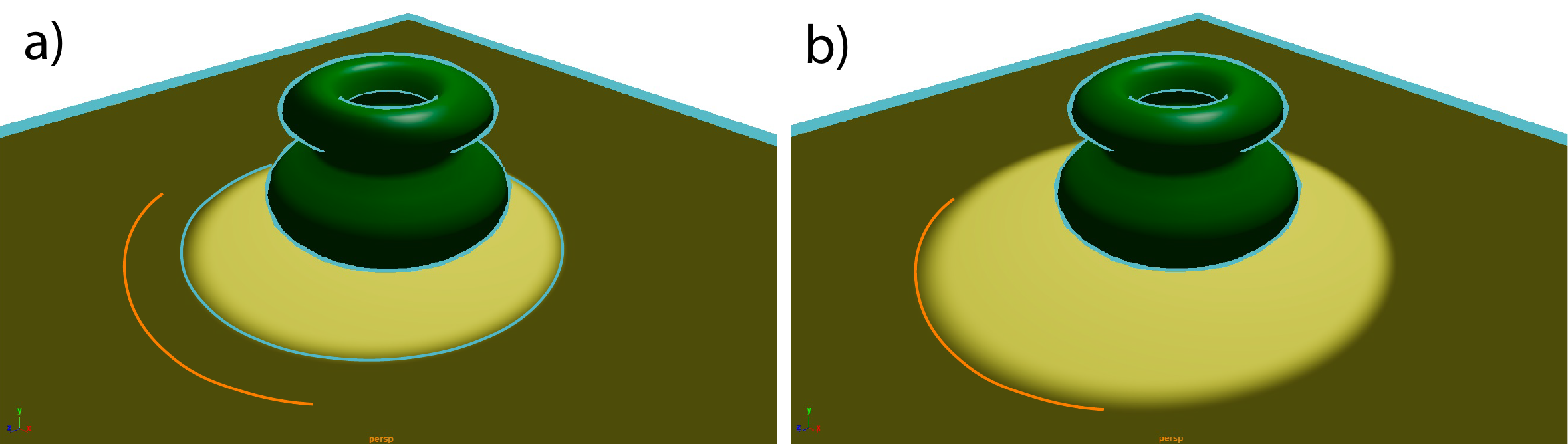}
    \caption{NPR and toon shading is used to compute \abscurve{abstraction curves} automatically for a 3D scene (a). User strokes can then manipulate scene attributes associated with these curves, like the cone angle of a scene spotlight (b). }
    \Description{A 3D scene demonstrating non-photorealistic rendering and toon shading.  a) shows an object within a spot light illuminating it and the user draws a curved arc in the dark area in the shape of part of the spotlight.  b) shows the spot light radius widening to match the curved arc the user drew.}
    \label{fig:abstraction}
    \vspace*{-1em}
\end{figure}

\addvspace{4pt}
{\bf Visual scene abstraction:}
The fundamental insight behind squidgets is that imagined curves in a scene provide natural manipulation handles. While such curves are often explicitly evident in 2D graphics (\autoref{fig:teaser}(f)), they need to be algorithmically inferred in 3D scenes (Figures \ref{fig:teaser}(a-e),\ref{fig:abstraction}).
Understanding and computing such a collection of curves that comprise a visual scene abstraction is addressed by research in non-photorealistic rendering
 \cite{gooch2001non} and perceptual psychology \cite{todd2004visual}. 
 We rely on this body of work to automatically compute a set of curves that define a visual scene abstraction comprising silhouette, ridge/ valley, shading contrast and border curves \cite{npr-tut}, as seen in \autoref{fig:abstraction}.

\addvspace{4pt}
{\bf Scene object transformation and composition:}
A large body of research spanning 50 years \cite{sketchpad} has specifically addressed the transformation (e.g. translation, rotation, scaling) of objects in 3D scenes. We refer readers to a recent survey paper \cite{3dmanip-star19} tracing this research from desktop to immersive devices. Techniques designed for constrained tasks like docking \cite{Bie90} have also been compared across various input modalities \cite{mousetouch3d-comp17}. Object manipulation techniques interact with a 3D spatial context that can be physical (such as a tangible 3D proxy), or virtual (like mouse controlled 3D widgets) \cite{bowmanbook04}. Squidget curves provide our spatial context, controlled by 2D (Figures \ref{fig:teaser}) or 3D (\autoref{fig:ax_cutter}) stroke input.

We note that transforming 3D objects with squidgets (\autoref{fig:teaser}) is but one example of our ability to homogeneously manipulate arbitrary scene object attributes using squidget strokes.

\addvspace{4pt}
{\bf Sketch-based gestures and interfaces:}
The recognition and use of stroke-based gestures can be categorized into gestures that are hard-coded or visually
matched \cite{gestures}. Hard-coded gestures tend to be expressive but context-specific and difficult to customize. Visual matching, such as the \$1 Gesture Recognizer \cite{dollar}, geometrically matches a stroke to a user-given set of gesture templates. These gestures are typically scene-agnostic and can be mapped to general directives (e.g. invoking undo using a scratch gesture).
Similar to visually-matched gestures, squidgets match user-drawn strokes to the set of squidget curves in a graphical scene.

User strokes, scribbles, and sketch gestures have been applied to interfaces in a variety of domains \cite{sketchmodeling-star2008,sketchcontent-star2020} including ideation \cite{arora2017sketchsoup}, modeling \cite{teddy,ilovesketch,mentalcanvas}, illustration \cite{Sykora09-EG,kitty,draco}, and animation \cite{lineofaction,Sykora05-SBM,hashim2023drawing}. Techniques to aid the execution of sketch strokes and to aid the drawing process have also been explored \cite{sketchrev}.
While our work is focused on an application-agnostic framework for general stroke-based scene manipulation, we draw inspiration from this body of work for compelling examples, such as an animated line-of-action \cite{lineofaction} in \autoref{fig:squidget-motivation}(b), \ref{fig:teaser}(h), to showcase squidgets interaction.

\addvspace{4pt}
{\bf Scene proxies and custom widgets:}
UI widgets are visual 2D/3D elements designed to provide an in-situ interface to manipulating objects and aspects of a virtual scene \cite{Bie86}. Sketching such UI sliders and axes for scene manipulation have also been explored~\cite{tsandilas2015sketchsliders, kwon2016axisketcher}.
Widgets are typically hand-designed to capture
the form/function of scene attributes they control, and can range from simple spatial transform widgets, to custom curve controllers used to illustrate mechanical assemblies \cite{mechanical} or specific to the deformation parameters of complex objects such as the human face in \autoref{fig:teaser}(g).
Point-click-and-drag interaction of such widgets can be improved using stroke-based techniques \cite{sketchwidgets08}.
Our squidgets framework is able to homogeneously interact using over-sketching with such curve-based widgets (\autoref{fig:teaser}(h)).

Curves have also been used as visual proxies and manipulation handles for deformable objects \cite{wires,silsketch}. 
Squidgets generalize this control beyond deformable objects (e.g. the avocado in \autoref{fig:teaser}(f)) to any scene elements whose visual appearance can be controlled by some set of attributes (e.g. the cone-angle of a spotlight in \autoref{fig:abstraction}).
Finally, sketched strokes have been anchored to objects in Augmented Reality scenes to provide dynamic visualizations of changing scene attributes \cite{realitysketch}, complementary to our problem where the scene attributes respond to the sketched strokes.

\addvspace{4pt}
{\bf Direct in-situ manipulation:}
In-situ visualization and control of object attributes in a scene can greatly streamline a sketch-based workflow \cite{xia-object}.
Squidgets take such a design further, allowing a user to directly sketch the visual change they expect as a result of changes in object attribute values. 

Such direct visual control requires an inverse mapping from the visual abstraction of objects and other scene elements to their attribute values. Inverse computation for direct control is popular in a number of domains on Computer Graphics, such as skeletal kinematics (IK) \cite{ik-survey,lineofaction}, facial expressions \cite{lewis2010direct, kim21}, CAD modeling \cite{DAGA21}, and animation \cite{spatkey,sketchimo}. Squidgets are inspired by these solutions from specific graphical contexts to build a general framework for manipulating sparse sets of scene attributes using sketch-strokes.

%% file: sections/3_squidgets.tex
\section{The Squidgets Framework}
\label{sec:squidgets}

\begin{figure*}[htbp]
    \centering \includegraphics[width=0.93\textwidth]{figs/squidgets_framework_basic_03.pdf}
    \caption{Squidgets enable manipulation of attributes that affect the visual appearance of a scene (a,b) through user selection $S_s$ and manipulation $S_m$ strokes (a single stroke can serve as both $S_s$, $S_m$). We infer the scene attributes $A$ a user wants to edit by finding a curve $C$ from among a set of scene curves that best match the stroke $S_s$ (c,d: select). We then infer the change of values from $A$ to $A'$  needed so the resulting curve $C'$ best fits the stroke $S_m$ (e,f,g: manipulate). }
    \Description{An illustration shows a scene where attributes are controlled through user-drawn strokes. The first set of frames (a,b) depicts the initial scene of objects with a chess piece and dice and their x,y,z coordinates.  The user draws a single selection stroke in (c) in the shape of the corner of the dice.  d) shows the abstraction curves of the chess piece and dice where the stroke is matching the corner abstract curve of the dice.  This selects the dice as the object to manipulate.  The user draws a manipulation stroke further away from the dice in the shape of the corner selected in e).  f) shows the change in x,y,z coordinates of the dice such that the die's abstraction curves fit the manipulation stroke.  This moves the dice to the manipulation stroke in g).}
    \label{fig:squidgets_basic_pipeline}
\end{figure*}

We introduce a stroke-based UI framework which develops the idea that any visible or perceived curves in a scene can serve as visual proxies and manipulation handles to scene attributes. Squidgets exploit these curves for rapid referencing and direct visual adjustment of scene attributes, illustrated in \autoref{fig:squidgets_basic_pipeline}: with the user workflow (top row) and underlying building blocks (bottom row). 
Within our framework, a squidget is a scene curve $C$ whose appearance is related to the values of an associated set of scene attributes $A$.

We naturally perceive a visual scene like \autoref{fig:squidgets_basic_pipeline}(a) as an abstracted collection of curves $\mathscr{C}$ like \autoref{fig:squidgets_basic_pipeline}(b). To make a change to the scene, users first draw a selection stroke $S_s$ in-situ to visually specify a curve $C \in \mathscr{C}$. For instance, the orange stroke in \autoref{fig:squidgets_basic_pipeline}c visually corresponds to the corner of the dice. Formally, given a selection stroke $S_s$
and an ensemble of scene curves $\mathscr{C}$, we need to compute a (partial) curve $C \in \mathscr{C}$ that best matches $S_s$ in a perceptual sense (Section~\ref{sec:curve}). 
Addressing the added complexity of partial curve matching is necessary: motivational examples, prior art \cite{silsketch}, and our experiments, all confirm that users naturally draw only as much of a scene curve as needed to convey a desired scene edit. For instance, only the part of the avocado silhouette that changes is drawn to convey a local deformation in \autoref{fig:teaser}(f).

Once $C$ is selected, users can draw a manipulation stroke $S_m$ to specify how $C$ should appear as a result of manipulating its associated scene attributes. In other words, the squidget curve $C$ serves as an interaction handle for its associated scene attributes. Formally, let $A$ be the set of scene attributes and $f$ the render function for a curve $C$, such that $C=f(A)$. The silhouette curve of the dice in \autoref{fig:squidgets_basic_pipeline}(b) for example, is a function of the dice shape, transform, and camera parameters. 

Now, given a user drawn manipulation stroke $S_m$, we need to find changed attribute values $A'$, such that the edited squidget curve $(C'=f(A')) \approx S_m$. For instance, modifying the translation attributes of the dice from $A$ to $A'$ in \autoref{fig:squidgets_basic_pipeline}(f) results in a curve $C'$ where the corner of the moved dice closely approximates the user stroke $S_m$. $A'$ thus represents the new set of attribute values which are applied to the scene, to result in the desired visual change to $C$ (\autoref{fig:squidgets_basic_pipeline}(g)). Note, that a single user stroke can functionally serve as both $S_s$ and $S_m$ (Section~\ref{sec:framework_manipulate}).

\subsection{Design Guidelines}
We distilled the following guidelines from multiple unstructured discussions with artists, observations of user scene mark-up, \add{previous experience with} in-situ manipulation workflows of scenes\add{, pilot-testing, and feedback over the iterations of our framework}.\\
\indent {\bf Curve Design:} All user-perceived curves in a scene are potential handles for squidgets interaction. These are either imagined visual \abscurve{abstraction curves} or curves explicitly modeled and rendered as scene objects. The explicitly modeled \rigcurve{rig curves} in a scene typically capture the construction history of objects, often meticulously hand-crafted by artists to provide in-situ control over scene object attributes. These artists expressed the need for an overall interface where such rig curves could operate homogeneously with abstraction curves, but where such curves (\bookmarkcurve{bookmark curves}) could be incrementally and efficiently created within the interface itself.\\
\indent {\bf Stroke semantics:} Users tend to draw simple and short strokes whenever possible to interact with the scene. Users often draw partial strokes with just enough context to unambiguously select a scene curve (e.g. only the corner of the eye shape in \autoref{fig:teaser}(g)), or to clearly convey the desired manipulation (e.g. only deformed part of the avocado silhouette in \autoref{fig:teaser}(f)). For small incremental edits, interaction can be streamlined by allowing a single stroke to serve as both the selection and manipulation stroke.\\
\indent {\bf Interactive manipulation:} Oversketched strokes typically convey rough edits. Finer control can be enabled by repeated oversketching, or by interactively dragging the stroke spatially.\\
\indent {\bf Attribute redundancy:} Complex scenes often have a large number of scene object attributes that can be manipulated to satisfy a desired manipulation stroke for \abscurve{abstraction curves}. The oversketched stroke in \autoref{fig:teaser}(b) for example, could be manifested by moving the lamp on the table as in \autoref{fig:teaser}(d), but also by moving the scene camera in the opposite direction, or by non-uniformly scaling the lamp to have a much thicker base. It is thus important to limit the set of attributes considered for abstraction curve manipulation, or have these associations explicitly defined with \rigcurve{rig curves} and \bookmarkcurve{bookmark curves}.

\subsection{Inferring Scene Curves and Attributes}
\label{sec:scene_properties}
Given a 2D/3D scene, existing NPR and toon rendering techniques \cite{npr-tut} allow us to efficiently and automatically create a collection of outline, silhouette, shading and feature poly-line \abscurve{abstraction curves} of scene objects (Figures \ref{fig:teaser}, \ref{fig:abstraction}). These curves comprise a scene abstraction $\mathscr{C}$. The attributes $A$ for any curve $C \in \mathscr{C}$ are typically those of its corresponding scene object. 
The rendering function $f$ specifies how these attributes $A$ define the visual appearance of $C$.  
For example, the control vertex positions of the avocado's outline curve and the spatial 2D transform attributes of the avocado object define its shape and location in the scene in \autoref{fig:teaser}(f). 
For shading discontinuities like the drop shadow in \autoref{fig:abstraction}, the attribute set expands from the objects casting the shadow to include the light attributes like the cone angle of a scene spotlight. 
Note that the set $A$ associated with any curve $C$ in our implementation is automatic and illustratively minimal.

\subsection{Stroke to Select a Squidget Curve}
\label{sec:framework_select}
User  \stroke{strokes} perform two functions: selecting a squidget curve (segment) based on similarity to a stroke; and changing scene attributes to make the selected squidget curve approximate a stroke. 

\begin{figure}[tbp]
     \centering \includegraphics[width=\linewidth]{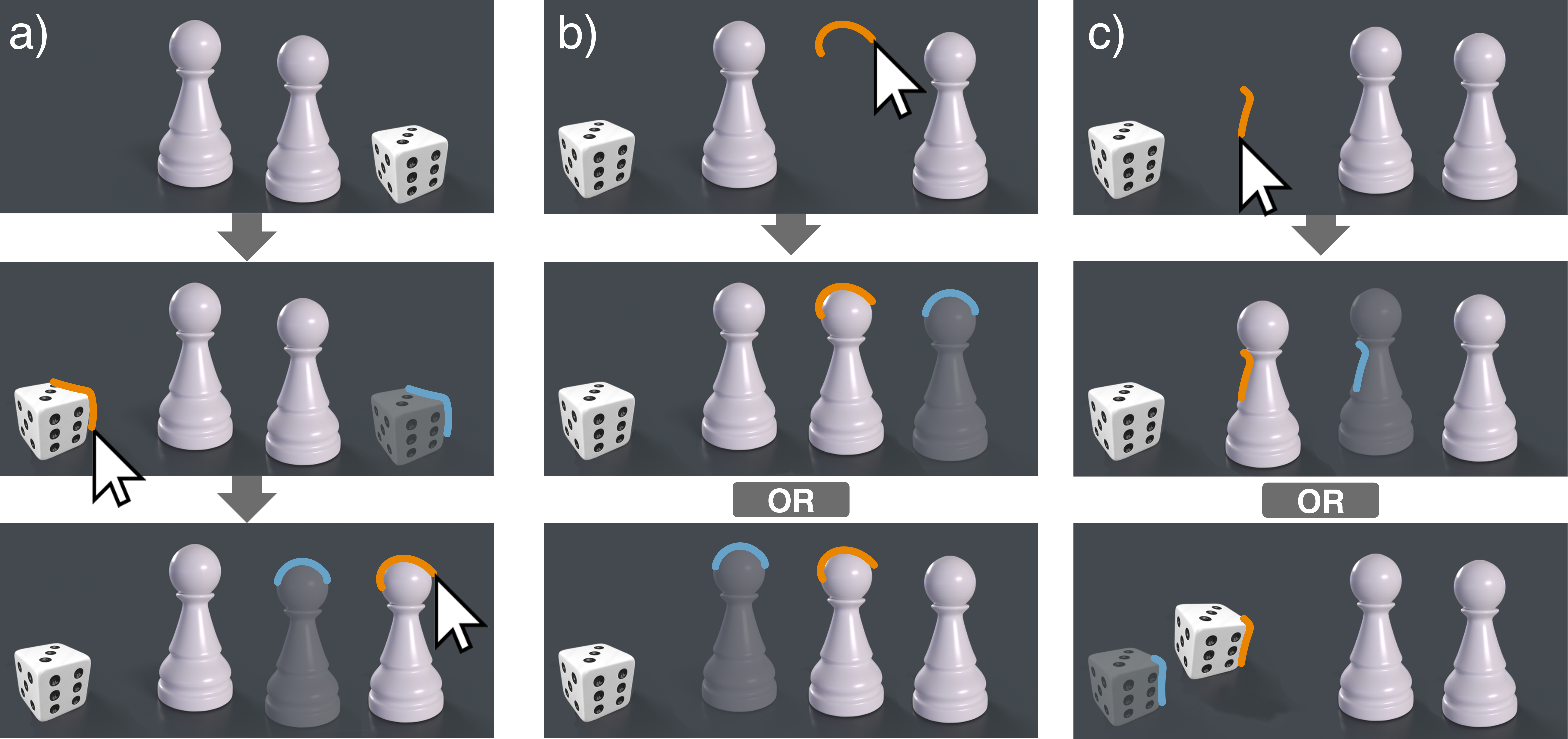}
     \caption{Disambiguation. Sketching unique shape features allows to unambiguously select and manipulate the dice in a one stroke interaction, even from a distance (a, middle). When multiple objects match the stroke, proximity is used for disambiguation (a, bottom). Ambiguities arise when two objects are an equally good match (b, c). In such cases, a two-stroke select-then-manipulate interaction is more suitable.}
     \Description{This figure shows 3 scenarios of how selecting an object first disambiguates object selection and manipulation.  a) shows an 2 of the same chess pieces and 1 dices.  By drawing the stroke distinctly to the shape of the chess pieces or dice, we can manipulate the intended object easily.  However, b) and c) show scenarioes where drawing a stroke could potentially select similar features in different objects.  Drawing 1 stroke to perform both selection and manipulation thus either both either of the 2 chess pieces (b) or either the chess piece or dice (c).}
     \label{fig:ambiguous_selection}
 \end{figure}

The similarity measure between a selection stroke $S_s$ and a target curve $C \in \mathscr{C}$ is highly dependent on scene context and application domain. For instance, in scenes where the curve shapes are distinct (e.g. rounded chess pawns vs. corners of a cubical dice in ~\autoref{fig:ambiguous_selection}(a), middle), a similarity measure based on shape alone can suffice. A shape-only metric however, performs poorly in cluttered scenes with many curve shapes being good matches for an input stroke (\autoref{fig:ambiguous_selection}(b,c)).
For example, in a scene with multiple chess pawns, a similarity measure that favors spatial proximity between $C$ and $S_s$ may be necessary to disambiguate selection (\autoref{fig:ambiguous_selection}(a), bottom). We present a novel perceptual curve similarity metric based on corner, spatial, and shape proximity in Section~\ref{sec:curve}. Note that such an algorithm can be further customized to add other matching criteria like stroke thickness and color, and additional heuristics constraining the attribute values search space.

The user's input stroke is matched for similarity against all curves in the scene, and the best matching curve is selected. Optionally, nothing can be selected if all matches are worse than a given threshold. Other mechanisms could also be leveraged for selected curve disambiguation, such as reducing the size of the curve set $\mathscr{C}$, e.g. using layer groupings of scene objects, or pre-selecting a specific scene object.

Note that a single point-and-click interaction is simply a degenerate case in our framework where the user selection stroke $S_s$ is a single point. Thus, a user can select the gaze-controlling rig curve in \autoref{fig:teaser}(g) with a point click (degenerate stroke) on or near any point of the eye-shaped curve, and then move the eye-shaped curve to a desired location by a subsequent click. Sketch strokes provide greater spatial and shape context for both curve selection and scene manipulation, than the spatial context provided by a single point. For example, the $<$ stroke drawn in different orientations can be used to both move and rotate the rig curve in \autoref{fig:teaser}(g).
Sketch strokes further enable the selection and manipulation of parts of the scene curves (\autoref{fig:teaser}(f)).

\subsection{Stroke as a Squidgets Manipulation Handle}
\label{sec:framework_manipulate}
Given a selected squidget curve, we assume the manipulation stroke $S_m$ is representative of its appearance as a result of associated scene attributes being changed from their current values of $A$ to $A'$ (in other words $C'=f(A')\approx S_m$).
This can be solved as an optimization where $A'$ is close to $A$ in value, and $f(A')$ is close to $ S_m$.
For the examples shown in our implementation, we are able to solve the above optimization as a best linear least squares minimization of transformation attributes and control point positions.

For small or incremental changes to a squidget's handle curve $C$, a single user stroke can specify both selection ($S_s$) and manipulation ($S_m$) (\autoref{fig:squidgets_basic_pipeline}). For stroke manipulation/deformation in complex scenes (e.g. \autoref{fig:ambiguous_selection}(b,c)), selection and manipulation steps may necessitate two strokes: an over-traced stroke to unambiguously select a desired squidget; and a subsequent manipulation stroke.

\subsection{Rig and Bookmark Squidget Curves}
\label{sec:framework_curves}
Artists increasingly hand-craft rig controllers: {\em in-situ} curve configurations pre-authored to control scene object attributes (\autoref{fig:teaser}(g)). Such \rigcurve{rig curves} are selected and manipulated homogeneously using strokes within our squidgets interaction framework. 

Scene manipulation and setup often entails interactive exploration, where key configurations of scene object attributes are incrementally discovered and need to be bookmarked for future use.
We support this workflow using \bookmarkcurve{bookmark curves}, drawn and explicitly associated with configurations of scene object attributes within the squidgets framework itself (\autoref{fig:teaser}(h)). Specifically, a user-drawn stroke can be associated with the current values of a set of scene attributes to define a discrete \bookmarkcurve{bookmark curve}. Selecting this curve using a squidgets framework at any point sets the associated scene attributes to their bookmarked values.

A number of such discrete \bookmarkcurve{bookmark curves} can also be chained to define a continuous piece-wise interpolation of the curves and their scene attribute values. Two discrete bookmark squidgets $C_0, A_0$ and $C_1,A_1$ for example, can be combined into a weight $w\in[0,1]$ interpolated, continuous squidget curve $interp(C_0,C_1,w)$ with similarly interpolated attributes $interp(A_0,A_1,w)$ for a set $A'$ that could be the union or intersection of attribute sets $A_0,A_1$.
Continuous \bookmarkcurve{bookmark curves} conceptually enable in-situ attribute keyframing (\autoref{fig:squidget-motivation}(a)),
visualized by a piece-wise linear path connecting a sequence of discrete \bookmarkcurve{bookmark curves}
(\autoref{fig:teaser}(h)).

Any object selection and curve modeling tool can be used to create \rigcurve{rig} and \bookmarkcurve{bookmark curves} with associated scene 
object attributes in 2D/3D.
Our implementation specially facilitates the drawing of \bookmarkcurve{bookmark curves} on planar canvases in 3D (\autoref{fig:scenario-squirrel}), where each canvas additionally
provides a grouping for \bookmarkcurve{bookmark curves} and their associate scene attributes.

%% file: sections/4_implement_framework.tex
\section{Implementation}
\label{sec:implement}

The squidget framework supports a range of implementations for a variety of applications and workflows. Here, we present details for an implementation\footnote{\add{\url{https://github.com/ohnooj/squidgets}}} within the 3D animation system {\em Maya} 2024. 
User stroke input is provided by a mouse, track-pad, or Wacom Cintiq 24HD Touch tablet.


\input{sections/curve}

\subsection{Squidget Curve Creation}
\begin{figure*}[!ht]
    \centering \includegraphics[width=\textwidth]{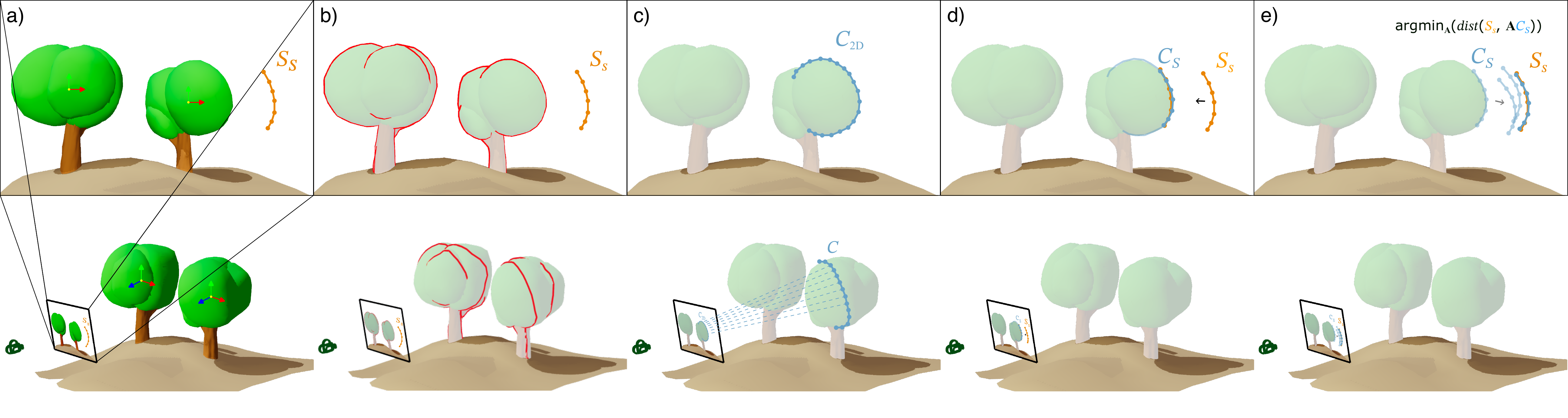}
    \caption{A user-drawn selection stroke is shown in screen space (top row, bottom row shows a world space view) (a). Maya toon render outlines are converted to abstraction curves in world space (b).  Each abstraction curve $C$ projected to screen space $C_{2D}$ (c) is aligned with $S$ (d), transforming $C_S$ to $S$ (e). The object corresponding to the best matching curve is selected for manipulation.}
    \Description{The top row shows the screen space view of 2 trees and the abstraction curve selection on a tree.  The bottom row shows the same scene in 3D.  From left to right: 1) 2 trees are there with a stroke to the right of the right tree indicating to move the tree, 2) the abstraction curves of the tree are generated, 3) One of the abstraction curves is chosen to match with the stroke, 4) the abstraction curve sub curve is found by projecting the stroke into the curve, 5) the curve is moved to the stroke through ICP.}
    \label{fig:abstool-select-impl}
\end{figure*}

While \rigcurve{rig curves} are user pre-authored 3D scene curves, \abscurve{abstraction} and \bookmarkcurve{bookmark} curves need to be created or set-up by our system.

\addvspace{4pt}
{\bf Abstraction Curves}

We use NPR toon-rendering in {\em Maya} to create imagined curves from scene objects (\autoref{fig:abstool-select-impl}(b)). Arbitrary object attributes can visually change these toon curves. Our implementation restricts these attributes to object transforms and object vertex positions for illustration. We further note that toon-rendering is view-dependent and toon curves can change discontinuously upon object or camera transformation. We thus create 3D curves $\mathscr{C}$ from 2D toon-rendered outlines given the current scene and camera parameters, as if they were pasted on their corresponding 3D objects, before any interactive scene manipulation begins.

\addvspace{4pt}
 {\bf Bookmark curves}
 
Like \rigcurve{rig curves}, \bookmarkcurve{bookmark curves} can be arbitrarily-created 2D/3D curves in {\em Maya}. Once created, they can be associated with any set of scene attributes within our framework.

{\em Bookmark canvases:}
We provide support for the streamlined creation and scene attribute association of bookmark curve groups using canvases.
 \add{
A canvas is a plane or surface that is positioned in 3D and associated with a set of scene attributes. 
Users draw a screen stroke projected onto the closest canvas to create an in-situ 3D bookmark-curve and automatically map the current values of the scene attributes to the curve.
Multiple squidgets can inherit the same scene attributes from the parent canvas but with different values such that groups of bookmark cruves can be managed via different canvases and spatially re-arranged in the scene by the user to reduce visual clutter. Visibility of canvases can also be toggled manually or automatically (e.g. based on proximity to a hovering pen position, or on how oblique a canvas is to the current scene view).
}
We experimented with other attribute selection schemes such as automatically selecting recently changed attributes, but found our proposed canvas-centric approach to provide a good balance between efficiency and flexibility in bookmark curve creation.


{\em Discrete/Continuous Bookmark Curves:}
\bookmarkcurve{Bookmark curves} can be linked in sequence by drawing a stroke that crosses discrete bookmark curves. This continuously interpolated bookmark is visually indicated by a poly-line path through the mid-points of discrete bookmark curves in sequence. The path further acts like a UI slider that provides linearly interpolated control between the scene attribute values associated with adjacent discrete bookmark curves.

\begin{figure*}[!ht]
    \centering
    \includegraphics[width=\textwidth]{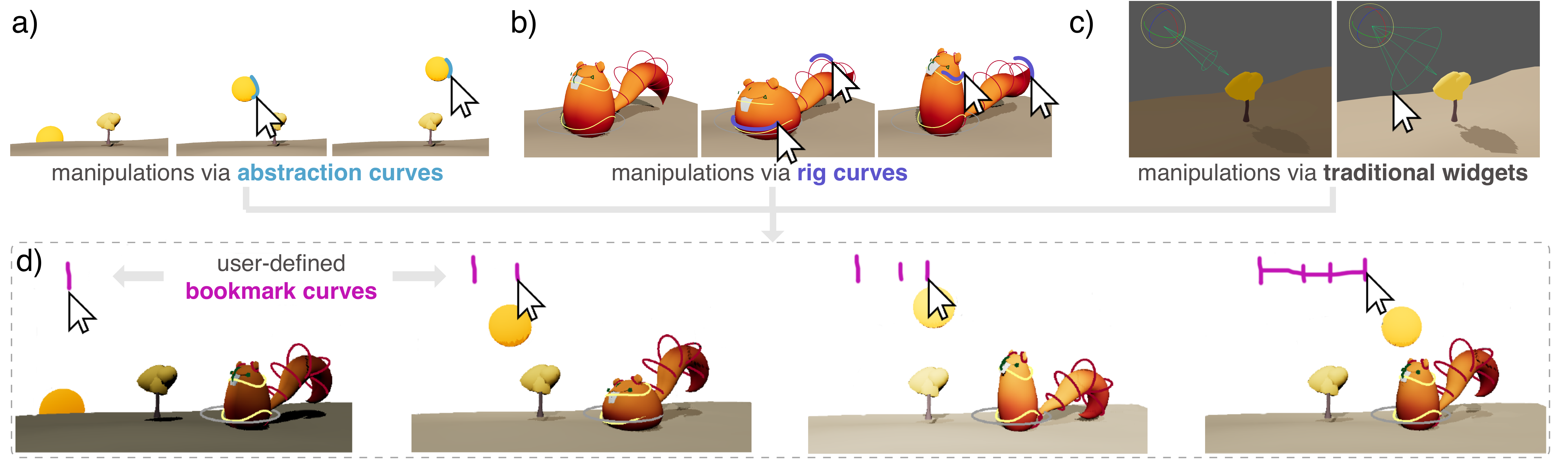}
    \caption{Top: To modify a scene, visual artists can freely choose between manipulating scene objects using \abscurve{abstraction curves} (a), \rigcurve{rig curves} (b), or traditional widgets (c). Bottom: The artist can save the scene (and hence all the scene parameters) by sketching a \bookmarkcurve{bookmark curve} (d, left). Iteratively modifying (a-c) and bookmarking scene configurations (d) allows the artist to create multiple versions, which can later be referred to, or interpolated by linking the bookmark curves (d, right).}
    \Description{This figure showcases how to use abstraction curves, rig curves, traditional widgets, and bookmark curves all together into 1 scene.  a) show how to manipulate the trajectory of a sun object with abstraction curves.  b) shows how to use rig curves to manipulate the handles of a squirrel rig to pose the squirrel character.  c) shows how to manipulate the illumination parameter of a light source using the traditional light widget.  d) shows how to use bookmark curves to key the 3 previous object's parameters into physical strokes.  By manipulating each of the 3 objects with their respective curves or widgets, we can create a sequence of bookmark curves to key multiple parameters.  We then draw a stroke to connect all the bookmark curves to create a continuous squidget for scene parameter interpolation.}
    \label{fig:scenario-squirrel}
\end{figure*}

\subsection{Squidget Curve Selection}
The curve matching algorithm (Section~\ref{sec:curve}) matches the selection stroke $S_s$ to the best matching 3D squidget (abstraction/ rig/ bookmark) curve $C$ projected into screen space \autoref{fig:abstool-select-impl}. Note that for 3D stroke input in AR/VR the matching takes place in 3D itself.

{\em Corner, Spatial, and Shape Weights}:
Our algorithm in Section \ref{sec:curve} allows us to differentially weight the importance of corners, a spatial transform, and the shape of curve segments in curve matching. A number of factors influence the weight settings, including: input device (trackpads and mice have greater noise than pen/tablet and thus a lower corner weight); separate select and manipulation strokes (the spatial weight for a dedicated selection stroke is low as we expect users to closely trace over the desired curve); nature of curves (for largely straight line segments in architectural applications the shape weight can be set as low, for known smooth curves the corner weight can be set low). \add{In a number of scenarios, assumptions about the scene and/or interactions allow to simplify aspects of the algorithm, as described in the supplemental materials.}

\add{We hand-tuned weights for application and stroke workflow (e.g. a 2-stroke workflow ($S_s \neq S_m$) expects an overtraced selection stroke and thus weights spatial proximity heavily). In scenes with many squidget curves, spatial proximity of the user strokes to the desired select/manipulate curves provides the most reliable behavior, given the corner and shape sensitivity of strokes to sketching inaccuracy.  Weights can also be learned using training examples of squidgets scenes and user strokes, or interactively set by a user.}


\subsection{Squidget Manipulation}
\label{sec:squidget_manipulate}
Once a squidget curve is selected, we need to compute the change to its associated scene attribute values, such that the resulting squidget curve best matches the manipulation stroke $S_m$. This search for optimal attribute values depends on the associated attributes. 
For example, for shape attributes like object vertex positions, the vertices can closely approximate the manipulation stroke by directly conforming object vertices to the shape of the manipulation stroke. For object transform attributes, there is an analytic best-fit transform that conforms a squidget curve to a manipulation stroke \cite{muller2005meshless}.
Discrete bookmark curves simply snap attributes to the bookmark curve closest to the manipulation stroke and continuous bookmarks are interpolated based on where the manipulation stroke intersects its poly-line path.
Abstraction and rig curves can have complex relationships between their associated scene attributes and the curve, requiring a neighborhood search of attribute values, to find the curve that best matches the manipulation stroke.

For interactive control of squidget curves, users can employ a 2-stroke technique where $S_s$ and $S_m$ are separate strokes, a 1-stroke technique where $S_s = S_m$, or a hold-and-drag technique to interactively refine the attribute values after drawing $S_m$, like a virtual slider that incrementally changes scene attributes or interactively changes the object transform.


It is important to note that when squidget curves are matched in screen space, all manipulations happen orthogonal to the camera.

%% file: sections/curve.tex
\subsection{Perceptual Curve Segment Matching}
\label{sec:curve}

\begin{figure}[htbp]
\centering \includegraphics[width=0.9\linewidth]{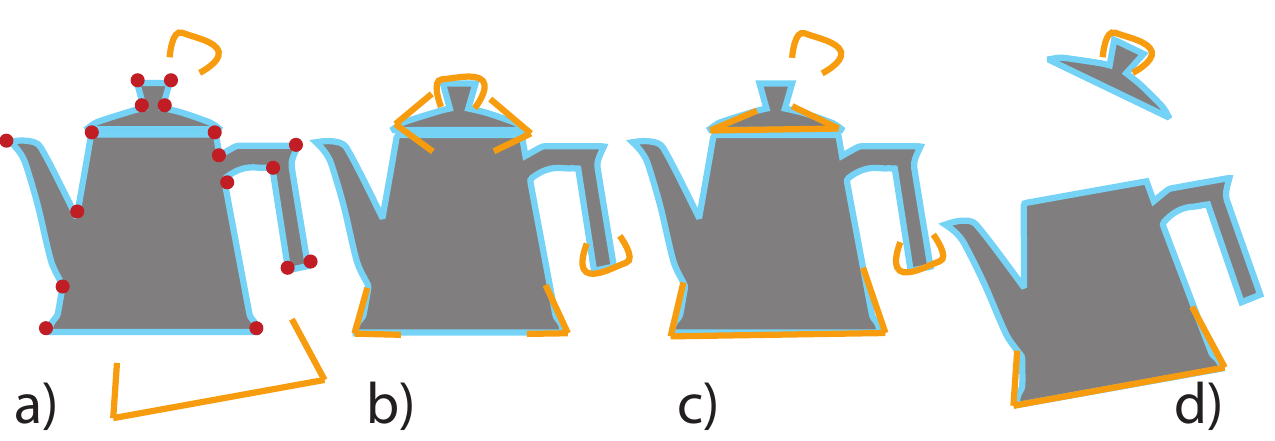}
\caption{Strokes (orange) being matched to curves (cyan) with corners (red) (a). Strokes can be partially matched and aligned by ICP, and corner matching (b). Curves segments can be parametrically matched (c) and the best matched curves used to manipulate scene attributes (d).}
\Description{This figure shows 4 panels of the stroke matching algorithm for a teapot.  Strokes are matched with corners of the teapot.  Strokes are aligned with the teapot by ICP.  Strokes and curves are parametrically matched.  The teaop lid is opened using the matching strokes.}
\label{fig:teapot}
\end{figure}

A general similarity metric for curve matching is an ill-posed problem, depending on its use case and assumptions on the geometric properties of the curves. 
Curve matching for most interactive graphics applications are based on {\em corner}, {\em shape}, and {\em spatial} similarity \cite{curvepartition}. Perceptually, we tend to align curves at corresponding sharp {\em corners} with matching smooth curve segments {\em shapes}. Depending on where the matching curves are drawn, {\em spatial} transforms may be needed to better align corresponding shape segments. We distill these observations into a curve matching algorithm with weights that favor different criteria, making the algorithm easy to adapt to specific use cases \add{(see details in supplemental)}, or fit to training curve data (\autoref{fig:teapot}).
We assume $P$ is an open curve for now and adapt our solution to closed curves later.

\vspace*{4pt}
{\bf Terminology}
\vspace*{-2pt}
\setlist[description]{font=\normalfont\itshape}
\begin{description}[labelindent=0.5em ,labelwidth=1em, labelsep*=0.5em, leftmargin =!, style = standard]%
\item[Curve:] we represent a curve $P$ as a poly-line sequence of points $P=<p_1,..,p_n>$.
\end{description}

\begin{description}[labelindent=0.5em ,labelwidth=1em, labelsep*=0.5em, leftmargin =!, style = standard]%
\item[Curve arc-length:] is computed, for a curve $P$, as the sum of poly-line segments $al(P,k)= \sum_{i=2}^k || p_i-p_{i-1}||$.
\item[Curve corners:] for a curve $P$, let $C(P)=<c_1,..,c_m>$, be indices for internal corner points $p_{c_i}$ in sequence. The corners thus induce smooth curve segments $P_l= <p_{c_l},.., p_{c_{l+1}}>$ (for $l \in \{1,m-1\})$, the end curve segments being $P_0= <p_{1},.., p_{c_{1}}>$ and $P_m= <p_{c_m},.., p_n>$.
Segment arc-length for an internal curve segments $P_l$ is thus $sal(P_l)= al(P,c_{l+1})-al(P,c_l)$. At the extremes $sal(P_0)= al(P,c_1)-al(P,1)$, and $sal(P_{m})= al(P,n)-al(P,c_{m})$.
\item[Curve corner position+orientation:] we define the position $o(P,i)= p_{c_i}$ for a corner point $i$ in the set of corners $C(P)$ of curve $P$. A coarse tangent-like orientation at this point is given by the vector $v$ between its adjacent corner points $t(P,i)= v(p_{c_{i+1}},p_{c_{i-1}})$, where $v(a,b)=(a-b)/||a-b||$. For the extreme curve segments we use $p_n$ as $p_{c_{m+1}}$ and $p_1$ as $p_{c_{0}}$.
\end{description}

{\bf Corner correspondence algorithm}

Given two curves $P$ and $Q$, we first find an optimal correspondence between the set of corners $C(P)$ and $C(Q)$,
that act as sparse anchors in perceptual shape matching \cite{curvepartition}. 
We use a dynamic programming formulation that minimizes the matching energy of corresponding corners on $P$ and $Q$.

Let $M(P,i,Q,j)$ be the {\em corner matching energy}  for the $i^{th}$ corner in $C(P)$ to correspond to the $j^{th}$ corner in $C(Q)$, defined as a weighted sum of a spatial alignment energy $A(i,j)$ needed to move and coarsely align the corners, and a coarse shape energy measured by a local disparity in curve arc-length of their adjacent curve segments. 
In other words:\\ 
\hspace*{4pt} $M(P,i,Q,j) = w_t * A(i,j) + w_s * ( | sal(P_{i-1}) - sal(Q_{j-1})| +\\ $
\hspace*{60pt} $ | sal(P_i)- sal(Q_j)| )$.\\
The corner alignment energy $A(i,j)$ above is a weighted sum of a translation $Tr(i,j) = o(P,i)- o(Q,j)$, and a rotation $Ro(i,j)$ that rotates vector $t(P,i)$ to $t(Q,j)$.

We need to adjust the above computation of coarse shape energy and corner alignment at end curve segments, to allow partial curve matching. For example, when $i=m$ let parameter $s=min(sal(P_m),sal(Q_j))$, and we use the point $q$ on $Q$ at parameter $al(Q,c_j) + s$ as corresponding to end point $p_n$ of $P$. The shape energy term above $| sal(P_i) - sal(Q_j)| $ is then replaced by $ | sal(P_i) - s| $, and the corner alignment $A(i,j)$ uses a translation $Tr(i,j) = o(P,i)- q$, and a rotation $Ro(i,j)$ that rotates vector $t(P,i)$ to $v(q,q_{c_j})$. The case for the beginning segment $P_0$, and the two end segments of $Q$ are treated similarly.


Now let $E_C(P,e,f,Q,g,h)$ be the energy to match a sequence of corners from $e,..,f$ in $C(P)$  to corners $g,..,h$ in $C(Q)$. We can define this energy using Dynamic Programming as:\\
\hspace*{4pt} $E_C(P,e,f,Q,g,h) = min_{i\in{e,f},j\in{g,h}}[ E_C(P,e,i-1, Q,g,j-1) + $ \\
\hspace*{80pt} $  E_C(P,i+1,f,Q,j+1,h) + M(P,i,Q,j)]$. 

{\bf Curve segment matching algorithm}

The matching corners provide anchors for a parametric correspondence between poly-line curve segments $P_l$ (for $l \in \{1,m\}$) and their matching segments in curve $Q$ (WLOG $|C(P)| \le |C(Q)|$). 
In the event that $P$ has no corners, we find a best-fit rigid transform that minimizes the distance between points on $P$ and the curve $Q$ using the iterative closest point ICP algorithm \cite{low2004linear}. We then find the closest pair of points on ICP aligned $P$ and  $Q$ and treat them as corners on $P$ and $Q$ respectively to provide an anchor for parametric correspondence. We also note that for the end segments $P_0$ and $P_{m}$, we use an arc-length parameterization to define a correspondence truncated by the segment of $P$ or $Q$ with a shorter arc-length. We now have two sets of parametrically matched curve segments in $P$, $Q$ induced by $P_l$ (for $l \in \{0,m\}$). Note again, for the case where $P$ has no corners we re-combine the closest-point-pair induced segments after parameter correspondence, so they are treated as a single smooth curve segment.
For each corresponding pair of segments $P_l$ and $Q_l$, we resample the sparser poly-line to match the point count $d$ of the denser segment. We now compare these segments simply using a matching set of corresponding points where $ P_l =\{pl_1,..,pl_d\}$ and $ Q_l =\{ql_1,..,ql_d\}$. The matching energy $M_l$ can be combined as a single best-fit rigid transform $A_l$ to align the point sets, and an average of residual distance-squared between corresponding points after the alignment transform. In other words,\\ \hspace*{10pt} $M_l= w_a * ||A_l|| + (w_p /d) * (\sum_{i=1}^d ||pl_i- ql_i||^2)$.\\ 
The overall curve matching energy $E(P,Q)$ between curves $P$ and $Q$ is a sum of the matching energies  of their corresponding curve segments $\sum_{l=0}^m (M_l)$.

For a given application, the best match between $P$ and a set of curves $Q^1,..Q^k$ is then simply $min_{i \in \{1..k\}} (E(P,Q^i))$, subject to setting the weight parameters to provide control over the impact of corners, spatial translation, rotation, and the shapes of the curves being matched.

For closed curves, we first perform a rigid ICP transformation to roughly align them, We then run the stroke matching above with a high spatial weighting to only allow small local movements of curve segments to align corners and curve segments.


%% file: sections/5_applications.tex
\section{Applications}
\label{sec:applications}

\begin{figure}[t]
    \centering
    \includegraphics[width=\linewidth]{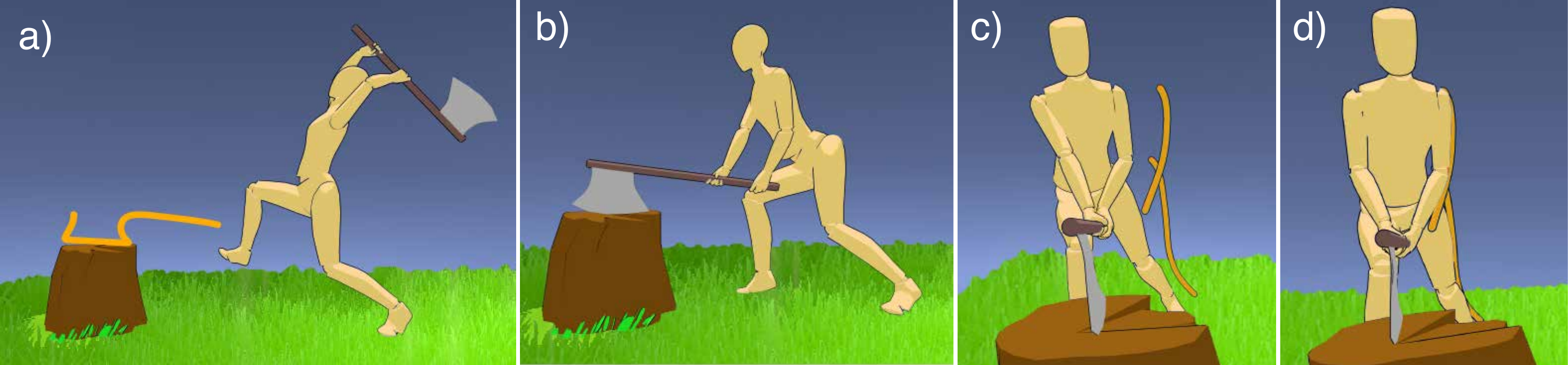}
    \caption{Within a VR environment, strokes are drawn to control a wood-chopper by moving the axe they hold (a-b) and adjusting the body pose (c-d).}
    \Description{A VR character is chopping wood.  We can use abstraction curves in VR to pose the character chopping wood and correcting the leaning of the character.}
    \label{fig:ax_cutter}
\end{figure}

\begin{figure}[t]
    \centering \includegraphics[width=0.95 \linewidth]{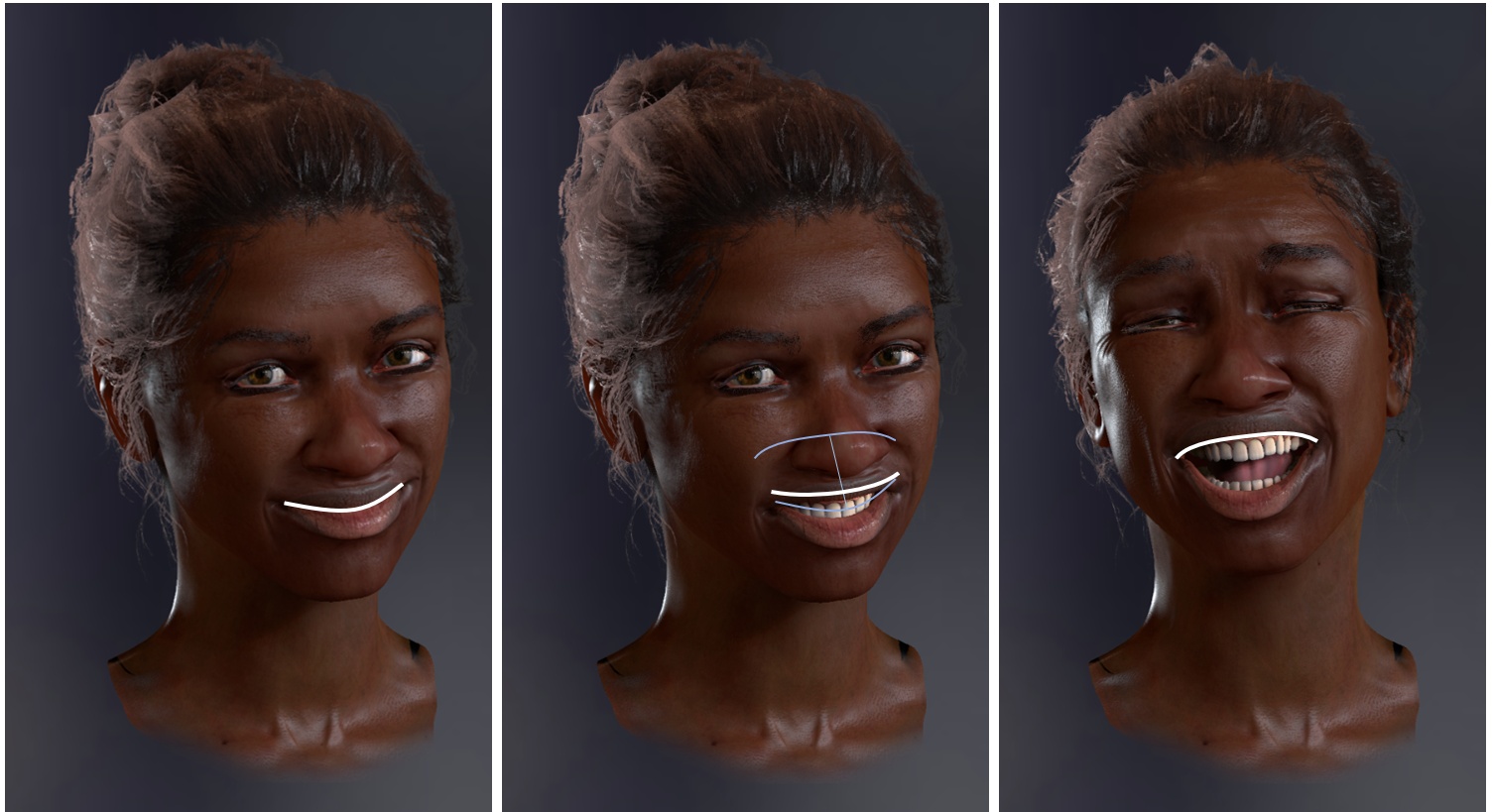}
    \caption{User-authored bookmark curves can be associated with discrete configurations of face attributes to capture a smile (left) and a laugh (right); and further combined into a continuous bookmark curve squidget, where the curves and related face attributes can be interpolated (center).}
    \Description{User-authored rig curves can be associated with discrete configurations of face attributes to capture a smile (left) and a laugh (right); and further combined into a bookmark curve squidget, where the curves and related face attributes can be interpolated (center).}
    \label{fig:lady}
\end{figure}

Most \textbf{2D design applications} already support sketch input, and can thus seamlessly incorporate squidgets for quick editing of scene parameters using \abscurve{abstraction curves}, as illustrated in \autoref{fig:teaser}(f). In \textbf{3D design applications}, in-situ control through squidgets can be a powerful approach compared to traditional widgets. For instance, in a \emph{VR environment} sketching strokes that match the scene's \abscurve{abstraction curves} can control characters, e.g. a stroke indicating where an axe should move \autoref{fig:ax_cutter}(a,b). Character and face rigs often have many rig curve handles for each part of the body, and posing often requires manipulating multiple handles together. Squidgets can facilitate \textit{character and face posing} by allowing artists to directly work with face (\autoref{fig:teaser}(g)) or character \rigcurve{rig curves} (\autoref{fig:scenario-squirrel}(b)), by drawing the shape of the rig handles. Using \bookmarkcurve{bookmark curves}, artists can also author their own "rig" curves, as illustrated in \autoref{fig:lady}.

Combining squidgets enables \textbf{rich authoring workflows}. \autoref{fig:scenario-squirrel} illustrates how an artist (a) manipulates the sun's location by drawing arc strokes to match the sun's \abscurve{abstraction curve}, (b) poses the squirrel by sketching strokes that match \rigcurve{rig curves} of its tail, head and body, and (c) changes the scene lighting using traditional widgets. When satisfied with the scene, (d, left) the artist creates a \bookmarkcurve{bookmark curve}, i.e. a simple tick mark for easy reference to this configuration. Then, (a-c) they continue to modify the scene, (d) adding a \bookmarkcurve{bookmark curve} for each new configuration. Through this workflow, the artist authors key frames of a short animation where the sun rises and the squirrel bounces, which they can interpolate by creating a continuous bookmark squidget (d, right).

%% file: sections/6_study.tex
\section{Evaluation}
\label{sec:study}

We evaluate our framework through a controlled experiment (\S\ref{sec:controlled_xp}), and an informal qualitative evaluation with visual artists (\S\ref{sec:xp_informal}). In the interest of space, we provide a summary; a thorough description and report can be found in the supplemental materials. The study was approved by our institutional ethics board.

\subsection{Controlled Experiment: Manipulate Objects}
\label{sec:controlled_xp}

We ran a within-subject experiment (n=12) to evaluate whether people understand the squidgets concept and the usability of using sketch input for manipulating scene parameters. We recruited 12 participants \add{who were familiar with pen input but with little experience using 3D design software} through our institutional channels and word of mouth. The study lasted $\sim$45min; participants received CAD\$30. The study ran on a computer laptop plugged into a Wacom Cintiq 24HD screen tablet; and was implemented with the Maya software with most features stripped out from the view to best approximate a generic, traditional vector graphics editor.

\subsubsection{Study Design}
Participants completed two sets of scene manipulation \texttt{Tasks}: \emph{translate} and \emph{deform} a graphical object to match a target. The \texttt{Techniques} included a \baseline with traditional graphical manipulation tools as found in vector graphics software \add{({\it Maya} selection/translation tool for translation tasks and single-vertex soft-selection with various deformation falloffs for deformation tasks)}, and different squidget variations: \onestroke (selection and manipulation are performed within a single input stroke), \selectdrag (selection with an input stroke, upon which the object sticks to the cursor to be further dragged around), and \selectmanipulate (an initial stroke is used for selection; subsequent drawn strokes perform manipulations to the selected object).


For the translation task, we varied difficulty according to the ambiguity that additional distractors would introduce for selection or manipulation: \texttt{SpatialDifficulty} refers to how far each distractor image is from the target axis location; \texttt{ShapeDifficulty} determines how visually similar in shape the distractor is to the task image. For the deform task, \texttt{DeformDifficulty} modeled task difficulty as a function of type of applied deformations. See supplemental material for details on the study design and procedure.

\subsubsection{Study Procedure} (1) The experimenter first obtained consent, then explained to participants that they were tasked with helping design a summer drink poster by moving or changing the shape of fruit stickers, using different techniques. (2) Because the study did not evaluate discovery but rather proficiency with the tool, the participants were allowed to practice for as long as desired with each \texttt{Technique} $\times$ \texttt{Task}. (3) Trial repetitions within each \texttt{Task} $\times$ \texttt{Technique} block were presented in a randomized order. For each trial, \texttt{Time}, \texttt{Operation}, and \texttt{Undo} were collected. After each block, participants were instructed to complete a NASA-TLX questionnaire~\cite{hart1988development}. (4) Finally, participants were invited to fill out a post-study evaluation asking them to rank the interactive techniques for each task, along with open comments.

\begin{figure}[tbh]
     \centering \includegraphics[width=\linewidth]{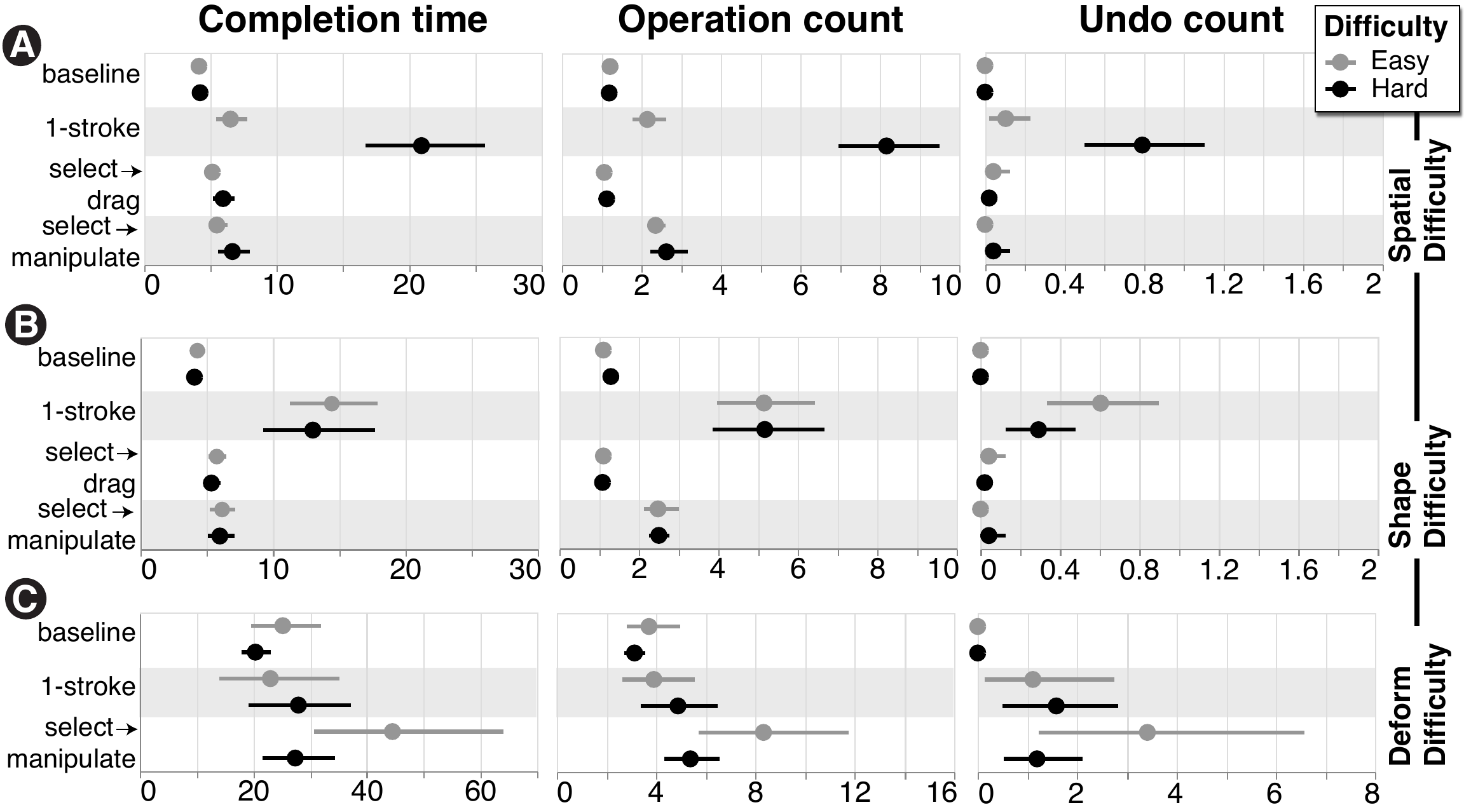}
     \caption{Effect of \texttt{Technique} and \texttt{Difficulty} on measures for the \texttt{translation} (A-B) and \texttt{deform} (C) tasks.}
     \Description{The figure shows the effects of Technique and Task difficulty on translation and deform tasks.}
     \label{fig:perf_all}
 \end{figure}

\subsubsection{Results}
We report the 95\% bootstrapped confidence intervals (CI) on means to assess effects~\cite{dragicevic2016fair} along with qualitative data. We focus on key insights; see supplemental for a detailed report.

\textbf{Performance: squidgets are comparable to the baseline most of the time.}
\autoref{fig:perf_all} shows the effect of \texttt{Technique} and task \texttt{Difficulty} on performance. Overall, the baseline was generally comparable to squidgets for both completion time and operation counts, across tasks, with a few exceptions, discussed below.

\begin{figure}[tbp]
     \centering \includegraphics[width=\linewidth]{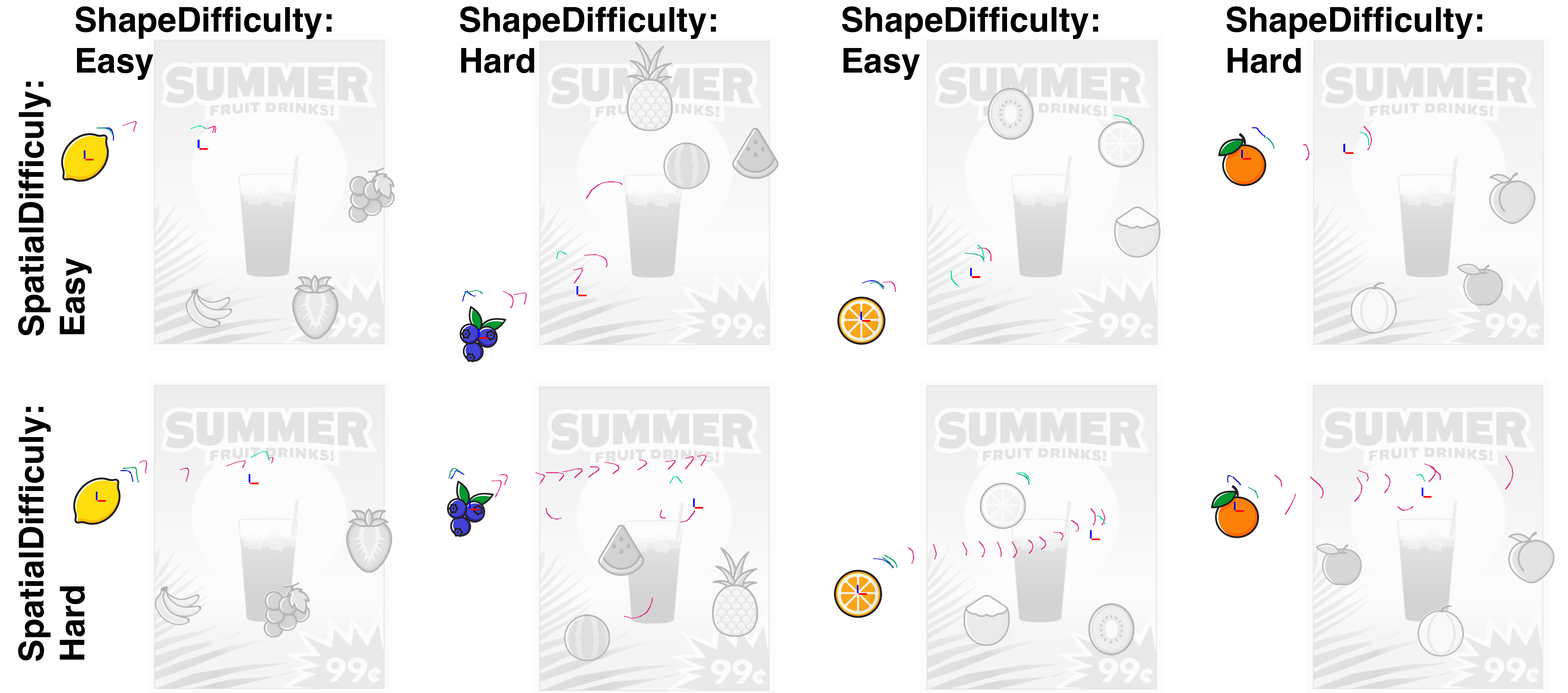}
     \caption{Move Task: P7's sketch strokes for \textcolor{c_1stroke}{\emph{1-stroke}}, \textcolor{c_selectdrag}{\emph{select$\rightarrow$drag}}, and \textcolor{c_selectpre}{\emph{select$\rightarrow$\textcolor{c_selectmanip}{manipulate}}}. \add{More in supplemental.}}
     \Description{Move Task: P7 sketch strokes for 1-stroke, select->drag, and select->manipulate for each of the 8 translation tasks.  Spatial difficulty is split between top (easy) and bottom (hard) while Shape difficulty is split between left (easy) and right (hard).}
     \label{fig:translation_overlay}
 \end{figure}

\textbf{For translations, distractors challenge \onestroke.}
As we anticipated, we find strong evidence that \onestroke presents notable challenges for move tasks with a \emph{hard} \texttt{SpatialDifficulty}. The strokes the participants drew (\autoref{fig:translation_overlay}) are mostly arc-shaped strokes that match a small portion of the objects' silhouette. While this strategy largely works when strokes serve one of two functions (selection or manipulation), for \onestroke there is increased challenge in the presence of distractors. When distractors were \add{spatially} close to the target, participants drew "incremental" simple strokes to move the object progressively \add{(\autoref{fig:translation_overlay}, bottom row)}. They did not systematically attempt to fine-tune the stroke to incorporate distinguishable features for disambiguation. Participants rated \onestroke as more mentally and physically demanding, more effortful, more frustrating, and less performing than the other techniques for moving tasks. However, two participants rated it as the most preferred, because \emph{``being able to do things in one stroke was overall much smoother compared to selecting.''} (P3).

\begin{figure}[tbp]
    \centering \includegraphics[width=\linewidth]{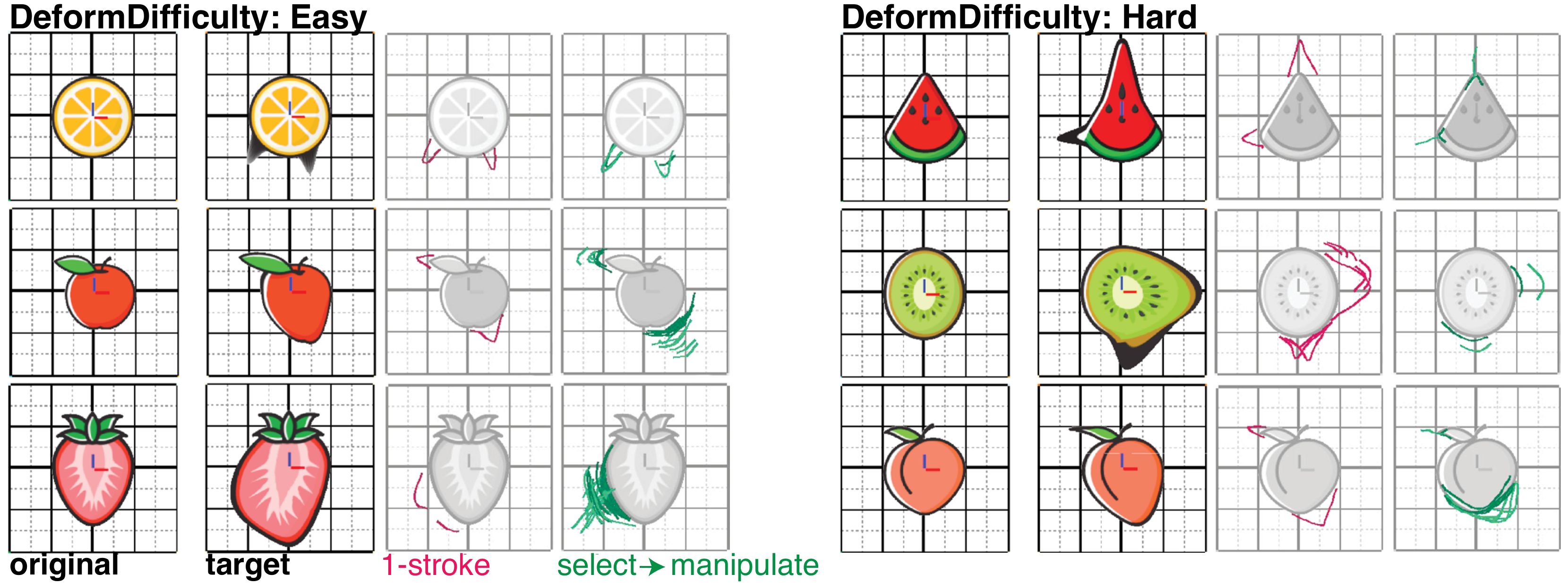}
    \caption{Deform Task: P7's sketched strokes for \textcolor{c_1stroke}{\emph{1-stroke}}, and \textcolor{c_selectpre}{\emph{select$\rightarrow$\textcolor{c_selectmanip}{manipulate}}}. \add{See supplemental for more.}}
    \Description{Deform Task: P7 sketch strokes for 1-stroke and select->manipulate for each of the 6 deformation tasks.}
    \label{fig:deformation_overlay}
 \end{figure}


\textbf{For translate tasks, different techniques will suit different people.} The best technique for move tasks depends on personal preferences. The most popular, \selectdrag (7/12 rated as most preferred), was liked for its efficiency  --- \emph{``The fastest and the most intuitive, and also had very good control in where you want to move the object''} (P2), and continuous interaction --- \emph{``You get to see real-time where the object is.''} (P10). Those who preferred the baseline (5/12) also mentioned high precision and control. Participants appreciated both techniques \emph{``were very simple and easy to understand right away.''} (P9). They were more mixed about \selectmanipulate, pointing to downsides such as \emph{``having the 2 stroke was very redundant compared to the other options''} (P8). While separating the select and manipulation strokes offers advantages conceptually, participants did not recognize these benefits during the controlled experiment.

\textbf{For deform tasks, a tool (baseline) is used iteratively while sketching the target (squidgets) is hit-or-miss.} The \baseline resulted in very few (if any) undo operations, but required about as many operations as \onestroke and \selectmanipulate. This indicates that, with the \baseline, participants opted for \emph{subsequent} manipulations instead of reversing the last operation. For squidgets, undo operations were more common, suggesting that participants imagined that the target deformation could be achieved in one go: \emph{``The deform-by-stroke tool feels more natural because that's how I anticipate deformation in my head''} (P10). However, due to the challenges in drawing curves which perfectly match the intent, this required multiple attempts, especially for lay users with no particular expertise with digital drawing.

\textbf{Deformations types do not correlate with task difficulty.}  \texttt{DeformDifficulty} had no effect, or even a countereffect on performance measures, suggesting that our metric is a poor proxy for anticipated difficulty. We found that participants struggled most with trials for which they had a hard time identifying what was deformed and how (i.e. peach \& apple, \autoref{fig:deformation_overlay}). Note that this is an artificially introduced and extraneous step which was inevitable for a controlled experiment but would not occur in a real-world scenario, where people know what target shape they want.

\textbf{For deform tasks, the sense of control is affected by the predictability of the technique and the dexterity with input.} While some participants found that the baseline offers more control and predictability, others felt the opposite was true because vertex-based distortion was constraining: \emph{``[the baseline] is my least favorite because you have to evaluate which width to use.''} (P10), and having poor predictability: \emph{``it's hard to map a larger deformation of the image to just a single vertex movement in my head.''} (P11). Similarly, some found \onestroke offers control and is highly predictable, whereas others found it did not behave as expected, e.g. \emph{``it moved parts that I did not necessarily want moved''} (P8). Opinions were also split for \selectmanipulate, with several finding that it gave \emph{``finer control''} (P10), but a few finding it was difficult to understand.

\textbf{For deform tasks, \selectmanipulate has strong potential.}  \selectmanipulate was a popular choice for deform tasks. One of its main advantages is fine control over the portion of the stroke concerned with the manipulation, which several participants noted was powerful. 
P11 articulated it best: \emph{``I like it best because I feel like it provided the most amount of control - I can control the length of the curve that I want to deform as well as how much I want to deform it.''}.

\subsection{Impressions From Visual Artists}
\label{sec:xp_informal}

\begin{figure}[t]
    \centering
    \includegraphics[width=\linewidth]{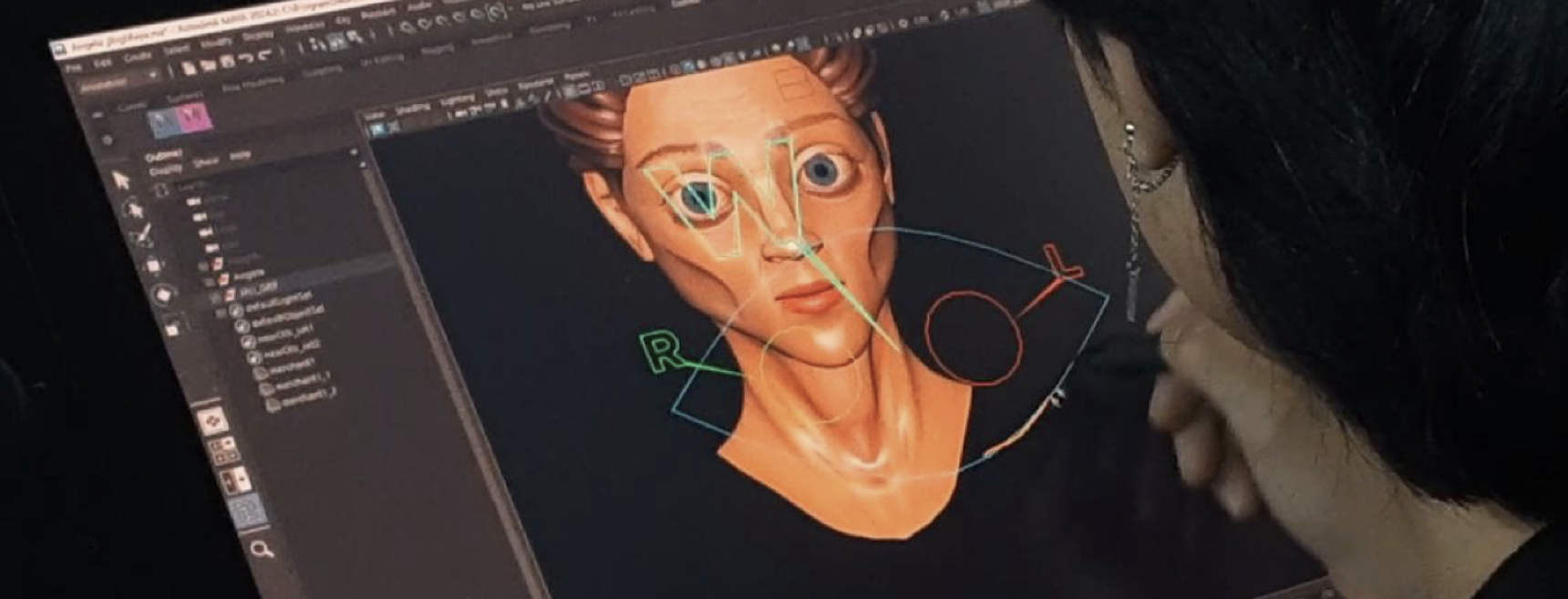}
    \vspace*{-15pt}
    \caption{An animator uses rig curves to control a character.}
    \Description{A photograph of an invited animator tester manipulating a character's gaze with squidgets on a large Wacom display tablet.}
     \vspace*{-10pt}
    \label{fig:animators}
\end{figure}

We invited \add{two} animators for interactive sessions to gather their impressions of the squidget framework on 3D scenes (\autoref{fig:animators}) in a formal 1-hour session where they explored all variations of squidgets while thinking aloud, followed by a semi-structured interview to discuss their experiences. A1 is a professional animator with $\sim$25 years of experience in the commercial and visual special effects industry, and whose primary tool is Maya. A2 is a researcher in graphics with 2 years of experience doing facial animation on Faceware\footnote{https://facewaretech.com/} and Maya. \footnote{\add{Four other expert animators also freely explored squidgets in earlier informal sessions, which allowed us to refine our framework.}}

\textbf{Conceptually, squidgets show promise.} Animators found that squidgets were well suited for creative workflows, qualifying the techniques as "very useful", "intuitive" and "very neat". They found sketch input \emph{in-situ} is a powerful mechanism that \emph{``brings everything to the center of the scene''} (A2), which helps maintain the creative flow. This contrasts with traditional interfaces: \emph{``as soon as you go into the interface, you're breaking what you're doing''} (A1). Animators found it to be \emph{``very satisfying''} when the system correctly interpreted their intent from \emph{``just putting a stroke down''} (A2). They showed excitement for the conceptual approach: \emph{``a software that captures your strokes, and eventually learn to predict what you're trying to do [...] That's a perfect piece of software''} (A1), albeit currently, the implementation needs improvements (see below). 

\textbf{Projecting a 2D stroke in a 3D scene.}
In our controlled experiment, we opted for 2D scenes due to the difficulties of 3D navigation, especially for lay users. Animators worked with 3D scenes, allowing us to gain insight into challenges and mitigating approaches. Our implementation interprets 3D transformations with respect to the camera view, which animators found \emph{``mentally confusing''} (A1) and misaligned with their expectations. They suggested alternative constraints to guide the transformations, such as axis constraints (x,y,z planar directions), or physics-based heuristics that result in \emph{``plausible movements''} (A2) like moving along a wall or undulating on a water body. A1 indicated that \emph{``even the stuff that is a little bit extraordinary is based on real world physics''}. Other scene objects were also indicated as useful guides to sketch over, for instance for assembly models, where sketching over objects will allow things to fit perfectly together \textit{"like a Tetris" (A1).}

\textbf{Disambiguation.}
Animators commented on disambiguation challenges exacerbated by near-by or too similarly shaped squidgets. They strove to balance the effort required to draw enough details or unique features for disambiguation. \selectmanipulate and \selectdrag achieved this goal successfully. One strategy we observed aimed to optimize for selection, then use minimal strokes for translations: \emph{``I tried to draw over the object; and then, it would just be like dashes. Very quick ones. [...] I like how even if you were lazy and don't want to sketch the full curve, you can also get there''} (A2). Animators also strategically picked parts that are simple to replicate to achieve precision in a two-stroke workflow. The squidget paradigm acts \emph{``differently than how you've selected for, you know, 30 years''} (A1), which can hinder adoptability. But animators indicated that sketch selection would work well to select distant objects when these are visually distinct.

\textbf{Squidgets workflow \#1: Coarse and granular changes.}
A1 discussed combining squidgets for coarse and granular changes, noting that \onestroke and \selectmanipulate are efficient for large movements but lack granular precision, while \selectdrag affords precision but is laborious for large manipulations. They described an envisioned workflow where \onestroke or \selectmanipulate interaction would first \emph{``snap the object in position''}, followed by refinements by dragging while swapping between translation and rotation to \emph{``just slightly move it exactly where it needs to be''}. 

\textbf{Squidgets workflow \#2: Pose and bookmark in-situ.}
A1 and A2 quickly picked up on the potential of squidgets for animation. They experimented with posing a character \cite{lineofaction} or a face using \abscurve{abstract} or \rigcurve{rig curves}, keying that pose with a \bookmarkcurve{bookmark curve}, repeating the operation multiple times with a new pose and key. Then, they linked the bookmark curves and scrubbed through the interpolated animation using \selectdrag. Within this workflow, the ease with which they could "re-key" or add a new pose without having to switch between the scene and a timeline would \textit{``save artists hundreds of hours''} (A1). A2 appreciated how the connected bookmark curves \emph{``visualizes the timing in the spatial domain''} akin to notations that artists use to indicate durations between frames; also reminiscent of timelines based on motion flow~\cite{dragicevic2008video, santosa2013direct, karrer2008dragon}.

\textbf{Bookmark curves as a personalized language.}
Animators quickly appropriated \bookmarkcurve{bookmark curves} as a personally-defined visual language. When bookmarking different character poses A1 and A2 drew the lines of action ~\cite{lineofaction} \emph{``following the spine and the legs of the character''} (A1), but later adopted minimal lines, as if the pose of the character was not necessary to fully ``encode''. Indeed, \emph{``making curves simpler will make it easier to query [...] I think just having the position is enough for me to specify which frame I want''} (A2). They also mentioned that custom markers would be useful for objects whose absolute position does not change, like faces.


\textbf{Scenarios of usage.} Concrete scenarios where animators see squidgets as a valuable alternative to their current practice include modularizing keyframe animation by creating sequences for smaller modules and controlling each squidget module with a meta-squidget; and the creation of \bookmarkcurve{bookmark curves} for intangible, difficult to reach scene parameters (e.g. lighting). A2 also felt that \selectmanipulate interaction with \abscurve{abstraction curves} would be particularly suitable for exaggeration effects, like squashing and squeezing.

\textbf{Areas for improvement.}
Squidgets have the potential to cause visual clutter if too many are created. Finally, because multiple different squidgets can be coupled to the same scene attributes, changing one squidget may affect the states of the others, causing concerns that such cases could yield undesirable results.

%% file: sections/7_conclusion.tex
\section{Discussion}
\label{sec:discussion}

Our prototype implementation and user studies affirm the feasibility of squidgets interaction and that the concept of squidgets, as a powerful and natural mechanism to express scene edits, was understood and appreciated by both experts and lay users. The studies confirmed our design guidelines and provided further insight into the squidget paradigm: (i) input device fidelity, drawing skill, and the choice of strokes for selection and manipulation, all affect the ability to select and perform desired complex scene edits; (ii) oversketching naturally aligns with 1-stroke interaction, but is only effective in simple scenes where disambiguating selection is easy, or when the desired edit is small (i.e. the selection and manipulation strokes are very similar); (iii) the \selectdrag workflow was appreciated for allowing a quick, coarse stroke for selection and manipulation that can be interactively refined by dragging;
(iv) an important difference between novice and expert users lied in their ability to draw uniquely identifying parts of strokes that aided selection and manipulation of scene objects. Experts also liked the gestural nature of squidgets interaction for quick scene exploration and bookmarking, and saw it as complementary to existing tools.


We present squidgets as a conceptual framework for interaction. Our implementation of squidgets within {\em Maya} is but one instance of a squidgets workflow. As our results show, there is potential within the framework to fine-tune the use of multiple strokes to disambiguate selection, vary the associated scene attributes, and the resolution and precision with which they are edited.



\subsection{Limitations}
\add{\textbf{Ambiguous inverse control:}} Conceptually, squidgets capture an ill-posed and ambiguous inverse control problem. Despite this, constraining the number of squidget curves and their mapping to scene attributes can make interacting with squidgets predictable and satisfying. 
There may be inherent redundancy in some scene attributes. For example, a stroke indicating a larger sphere in {\em Maya} can be realized by a change to the sphere's radius, its scale transform, or an infinite combination of the two. Such redundancy can be mitigated by regularization or authoring explicit bookmark curves that clarify the attributes to be controlled by the squidget.
Our prototype implementation also makes a number of simplifying assumptions such as only controlling transform and shape attributes via abstraction curve squidgets.
Given recent research on data-driven inverse rendering, we hope to solve general scene attribute manipulation for abstraction curve squidgets as an optimization learned from forward rendering simulations.

\addvspace{4pt}
\add{
\textbf{Confidence and predictability:}
The curve matching interaction may leave users with little transparency regarding output prediction and can leave users guessing what will happen.  Users commit to drawing an entire stroke before receiving feedback on their actions, and if the result is not what a user wants, there may be little guidance on how to draw a more accurate stroke for their desired task.
Several approaches can be explored for increasing confidence and predictability as follows.
}

\add{
For squidgets selection: The \onestroke workflow is best suited (1) when distinct stroke gestures are unambiguous, i.e., there are few, visually distinct squidget curves, or (2) for incremental changes, i.e., the difference between the selected and manipulated curve is small. While more efficient, this workflow can be unpredictable, especially in scenes with many squidgets. A correction step can be considered where the user picks among the top few visualized matches. The \selectmanipulate workflow is very predictable if the selection stroke traces over even a small portion of the desired curve (akin to click-to-select). Visualizing the best curve match to the partially drawn stroke in real time can increase user confidence.}

\add{For squidget manipulation: Dynamically rendering a ghost preview of the manipulated curve that best matches the user stroke can communicate how squidgets interprets input in terms of scene attribute changes. \selectdrag allows interactive scene refinement through spatial dragging; it does not allow refining the stroke shape though, something which could be considered using oversketching before committing a scene edit. Finally, a squidgets interaction history can also be used to reinforce repeated selection and manipulation behaviors, and penalize those that are undone.
}


\addvspace{4pt}
\add{
 \textbf{2D input:}}
While our framework supports 2D/3D scene curve manipulation, users interact with 2D strokes drawn in screen space (with the exception of mid-air strokes drawn in AR/VR). As a result, scene attributes are edited to produce view-depth preserving edits of the selected scene curve. The onus is thus on a user to pick good views to draw strokes, avoiding views where desired attribute edits significantly impact the scene depth.

\addvspace{4pt}
\add{
 \textbf{Deformation difficulty:}
We chose simple bell-shaped deformation tasks for fair comparison between the proportional modification baseline tool in Maya and our squidget implementation. We acknowledge though that squidgets allow more expressive shapes with any number of waves and swirls, than does our baseline. Future work should investigate alternative measures that more closely align with difficulty prediction.
}


\subsection{Future Work}
\textbf{Multi-stroke input:} Squidgets, like single-stroke gestures, currently use a single input stroke for selection/manipulation. Given the ability to draw multiple strokes before they are processed for selection/manipulation can greatly increase the expressive power of squidget interaction. For example, a pair of parallel strokes could better select thin tubular objects (a wine glass stem), or three intersecting strokes could indicate a 3D frame of reference providing better control over attributes that change scene depth.

\addvspace{4pt}
\textbf{Stroke attributes:} 
Our experiment only focused on geometric shape matching for scene curve selection and manipulation. Sketch strokes themselves have a rich set of attributes like weight, color and style, that could be exploited to further control the selection and manipulation of scene curves and associated attributes. 

\addvspace{4pt}
\textbf{Bookmark curve extensions:} Many of the animators in \autoref{sec:xp_informal} saw the value of \bookmarkcurve{bookmark curves} to bootstrap and improve in-situ graphical interfaces. 
Bookmark curves, currently used to represent discrete scene configurations, or interpolated in sequence can be embedded in-situ on manifolds to allow a richer exploration space of scene attributes \cite{arora2017sketchsoup}. Bookmark curve templates that transfer curves and associated scene attributes across objects is also exciting.

\addvspace{4pt}
\add{
\textbf{3D Navigation:} A crucial element of graphics workflow is camera navigation.  Though we did not explicitly implement navigation methods (and left users to use default camera controls), squidgets lays a foundation to build upon stroke-based 3D camera navigation. \rigcurve{Rig curves} and \bookmarkcurve{bookmark curves} can be pre-authored like typical 3D navigation widgets, to allow camera navigation control, or for constrained 3D navigation to key 3D configurations or along transformation trajectories.  \abscurve{Abstraction curves} automatically inferred from a 3D scene-view are particularly effective for view-dependent on-screen navigation; curves can provide useful alignment and docking constraints in a scene to further aid rapid 3D navigation.  Further exploration would be required to understand user tendencies to explicitly convey camera actions with strokes.
}

\section{Conclusion}
\label{sec:conclusion}
Squidgets present a grand unified vision for stroke-based interaction that leverages human perception of images as a collection of real or imagined curves. Our prototype only implements a small subset of squidget interactions in 2D/3D scenes with 2D/3D stroke input. 

While our studies do also point out challenges in squidget interaction for scene attribute control, we believe these can be addressed by better constraining curve selection and attribute manipulation to the task context and application domain.
Squidget do place expectations on user drawing skill, but we believe these can be reduced by 
simple and unique squidget curve shapes, stroke beautification techniques, incremental refinement via multiple strokes, and visual feedback that guides and snaps users towards better strokes.

Squidgets are a novel interaction framework that couples our natural tendency to indicate changes to a visual scene by drawing strokes over it. We have shown using our implementation and studies that such an approach is both viable and promising and hope that this work will fuel further work in direct, in-situ, sketch-based scene interaction.
